\documentclass[twocolumn,english,aps,prb,superscriptaddress,letterpaper]{revtex4}
\pdfoutput=1
\usepackage{graphicx}
\usepackage{amssymb,amsmath}
\usepackage{times}
\usepackage{subfigure}
\usepackage{color}
\usepackage[colorlinks=true,linkcolor=blue,citecolor=red,urlcolor=black,pagebackref=true]{hyperref}
\usepackage{textcomp} 
\usepackage[bold]{hhtensor} 
\begin{document}
\title{Synchronization of Micromechanical Oscillators Using Light}

\author{Mian Zhang$^\ast$}
\affiliation{School of Electrical and Computer Engineering, Cornell University, Ithaca, New York 14853, USA.}

\author{Gustavo Wiederhecker$^\ast$}
\affiliation{School of Electrical and Computer Engineering, Cornell University, Ithaca, New York 14853, USA.}
\affiliation{Instituto de F\'{i}sica, Universidade Estadual de Campinas, 13083-970, Campinas, SP, Brazil.}

\author{Sasikanth Manipatruni}
\affiliation{School of Electrical and Computer Engineering, Cornell University, Ithaca, New York 14853, USA.}

\author{Arthur Barnard}
\affiliation{Laboratory of Atomic and Solid State Physics, Cornell University, Ithaca, New York 14853, USA.}

\author{Paul McEuen}
\affiliation{Laboratory of Atomic and Solid State Physics, Cornell University, Ithaca, New York 14853, USA.}
\affiliation{Kavli Institute at Cornell for Nanoscale Science, Ithaca, New York 14853, USA.}

\author{Michal Lipson$^{\ast\ast}$}
\affiliation{School of Electrical and Computer Engineering, Cornell University, Ithaca, New York 14853, USA.}
\affiliation{Kavli Institute at Cornell for Nanoscale Science, Ithaca, New York 14853, USA.}
\date{\today}

%% Notice placement of commas and superscripts and use of &
%% in the author list
\topmargin -1.5cm
\oddsidemargin -0.3in
\textwidth 18cm 
\textheight 23.2cm

\baselineskip12pt

\begin{abstract}
\begin{center}
\small{$^{\ast}$These authors contributed equally to this work.}\\
\small{$^{\ast\ast}$To whom correspondence should be addressed; E-mail:  ml292@cornell.edu}\\
\end{center}
Synchronization, the emergence of spontaneous order in coupled systems, is of fundamental importance in both physical and biological systems. We demonstrate the synchronization of two dissimilar silicon nitride micromechanical oscillators, that are spaced apart by a few hundred nanometers and are coupled through optical radiation field. The tunability of the optical coupling between the oscillators enables one to externally control the dynamics and switch between coupled and individual oscillation states. These results pave a path towards reconfigurable massive synchronized oscillator networks.
\end{abstract}
\maketitle

\date{\today}
\newcommand{\nocontentsline}[3]{}
\newcommand{\tocless}[2]{\bgroup\let\addcontentsline=\nocontentsline#1{#2}\egroup}

% !TEX root = sync_arXiv.tex      
Synchronization processes are part of our daily experiences as they occur widely in nature, for example in fireflies colonies \cite{BucBuc6803}, pacemaker cells in the heart \cite{Pes75}, nervous systems \cite{WhiTraJef9502} and circadian cycles \cite{Str03}. Synchronization is also of great technological interest since it provides the basis for timing and navigation \cite{Bah09}, signal processing \cite{Bre02}, microwave communication \cite{KakPufRip0509}, and could enable novel computing \cite{MahYam0805} and memory concepts\cite{BagPooLi1110}. At the micro and nanoscale, synchronization mechanisms have the potential to be integrated with current nanofabrication capabilities and to enable scaling up to network sizes. The ability to control and manipulate such networks would enable to put in practice nonlinear dynamic theories that explain the behaviour of synchronized networks \cite{Str98,HopIzh0102}. Recent work on coupled spin torque \cite{KakPufRip0509,ManRizEng0509} and nanoscale electromechanical oscillators (NEMS) \cite{ShiImbMoh0704,ZalAubPan0310} exhibit synchronized oscillation states. Major challenges with synchronized oscillators on the nanoscale are neighbourhood restriction and non-configurable coupling which limit the control, the footprint and possible topologies of complex oscillator networks \cite{HeiLudQia1107,MarHarGir0603,Mar1110}. Here, we demonstrate the synchronization of two dissimilar silicon nitride (Si$_3$N$_4$) self-sustaining optomechanical oscillators coupled only through the optical radiation field as opposed to coupling through a structural contact or electrostatic interaction \cite{BukRou0212,KarCroRou0904}. The tunability of the optical coupling between the oscillators enables one to externally control the dynamics and switch between coupled and individual oscillation states. These results pave a path towards realizing synchronized micromechanical oscillators systems connected through optical links.

\begin{figure}
\begin{center}
\includegraphics[scale=0.9,angle=0]{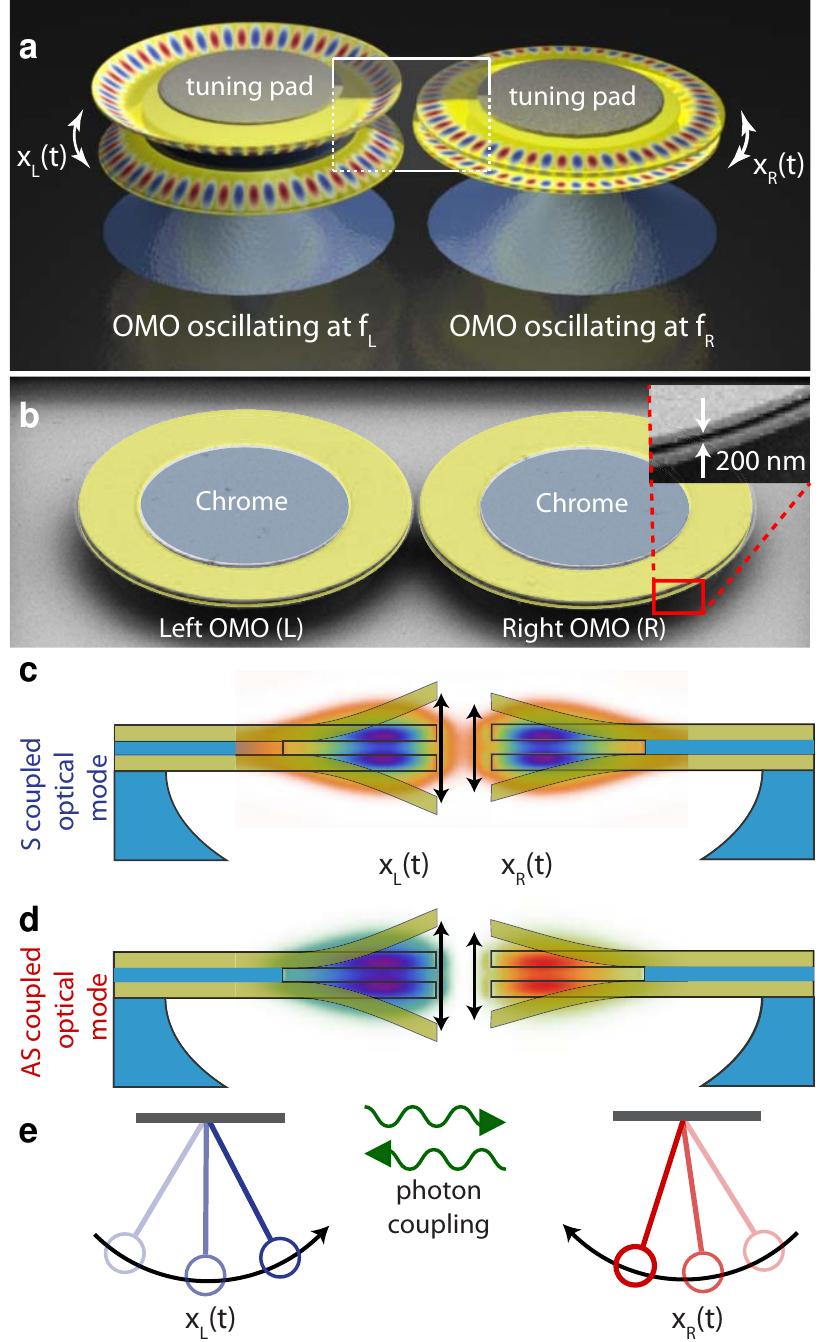}
\caption{Design of the optically coupled optoemchanical oscillators (OMOs).  (\textbf{a}) Schematic of the device illustrating the mechanical mode profile and the optical whispering gallery mode. (\textbf{b}) Scanning electron micrograph (SEM) image of the OMOs with chrome heating pads for optical tuning by top illumination. (\textbf{c},\textbf{d}) The symmetric (S) and anti-symmetric (AS) coupled optical supermodes. The deformation illustrates the mechanical mode that is excited by the optical field. (\textbf{e}) The dynamics of the coupled OMOs can be approximated by a lumped model for two optically coupled damped-driven nonlinear harmonic oscillators.}
\label{fig:fig1main}
\end{center}
\end{figure}

Optomechanical oscillators (OMOs) consist of cavity structures that support both tightly confined optical modes and long-living (high quality factor) mechanical modes \cite{KipVah07, BraStrVya0109}. These modes can be strongly coupled: the cavity optical field leads to optical forces acting on the mechanical structure; mechanical displacements due to this force in turn affect the cavity optical field. Amplification or cooling of the mechanical modes of these cavities can be achieved by feeding these cavities with a continuous-wave (CW) laser \cite{ArcCohBri06}. The mechanical vibration (driven by thermal Brownian motion) induces fluctuations of the cavity length, which translates into fluctuations of the optical resonant frequency; for a fixed-frequency driving laser this implies that the optical energy stored in the cavity also oscillates. Due to a finite cavity optical lifetime, the optical field does not respond instantaneously to the mechanical motion but instead oscillates with a slight phase lag; as the force that the optical field exerts on the mechanical mode is proportional to the stored optical energy, it will also experience such delay. Consequently the optical force will have one component that is in phase with the mechanical displacement, and another component that is $90^{\circ}$ out of phase. When the laser is blue (red) detuned with respect to the optical mode frequency, the optical force component that is in phase with the mechanical vibration induces an optical spring effect that increases (reduces) the mechanical frequency \cite{SheGraMow0405}, thus stiffening (softening) the mechanical spring. The out of phase component will decrease (increase) the effective mechanical damping, thus amplifying (cooling) the mechanical oscillation. Above a certain threshold laser power this optomechanical amplification overcomes the intrinsic mechanical damping; the device evolves from an optomechanical resonator to a self-sustaining optomechanical oscillator (OMO) \cite{MarHarGir0603}. The laser signal fraction that is transmitted, or reflected, from the optomechanical cavity becomes deeply modulated at the mechanical frequency of the oscillator \cite{CarRokYan05,KipVah07,GruLeePai1002}.

\begin{figure*}
\begin{center}
\includegraphics[width=0.9\textwidth,angle=0]{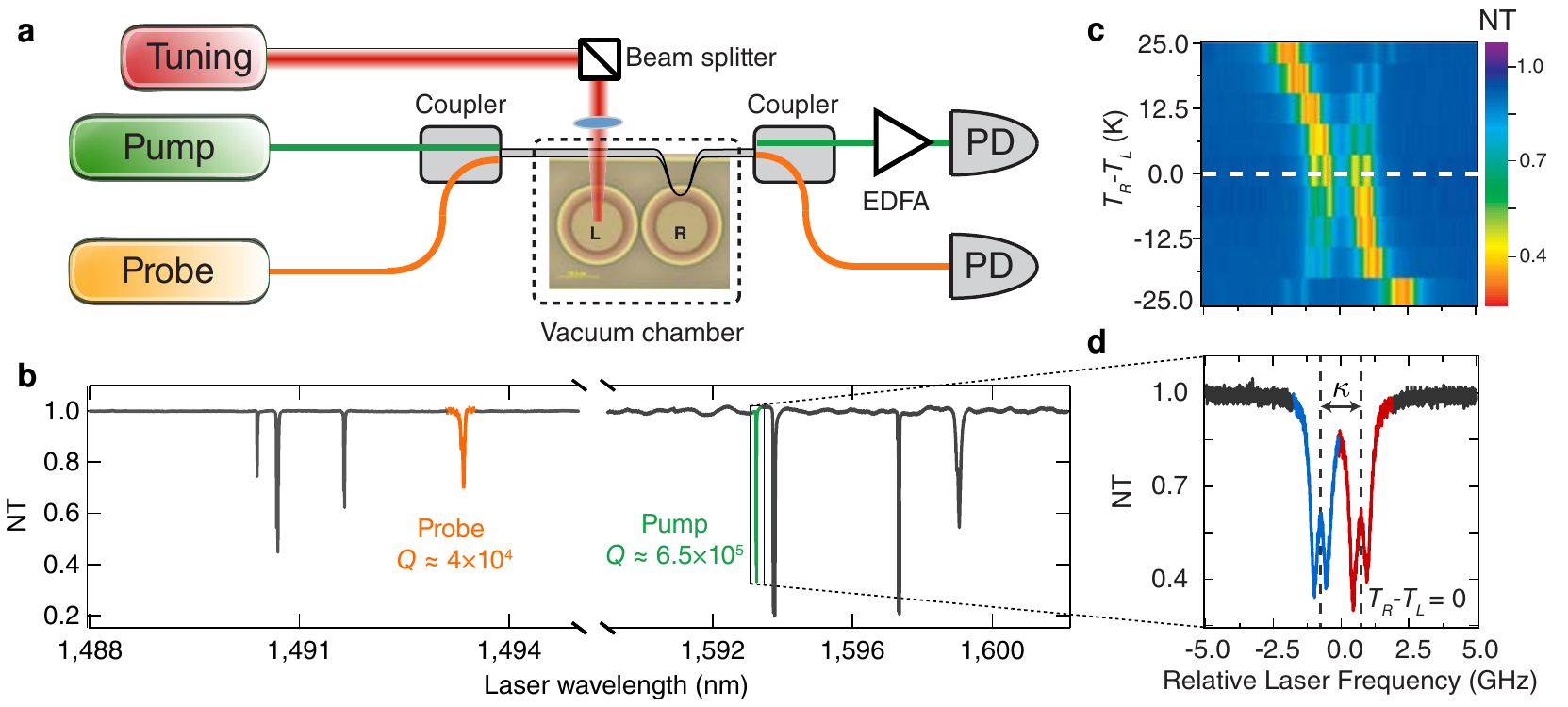}
\caption{Controlling the two OMO system. (\textbf{a}) Schematic of the experimental setup. The pump and probe light are launched together into the cavities and are detected separately by photodiodes (PD). An erbium doped fibre amplifier (EDFA) is used to amplify the transmitted signal to increase the signal strength. (\textbf{b}) Transmission spectrum of the coupled cavities. The green and orange coloured optical resonances correspond to the pump and probe resonances respectively. NT: normalized transmission. (\textbf{c}) Anti-crossing of the optical mode as the relative temperature of the $L$ OMO ($T_L$) and the $R$ OMO ($T_R$) is changed through varying the tuning laser power. The tuning laser is focused on to the two OMOs respectively to obtain the negative and positive relative temperatures. (\textbf{d}) Transmission spectrum of the maximally coupled state indicated by the white horizontal line in (c). The red (blue) part of the curve indicates the anti-symmetric (symmetric) optical supermode, $\kappa$ is the optical coupling rate.}
\label{fig:fig2main}
\end{center}
\end{figure*}

Recently it has been predicted that a pair of OMOs could synchronize if they are optically coupled as opposed to mechanically coupled \cite{ManWeiLip11,HolMeaMil1105}. Here we experimentally demonstrate the synchronization of two optically coupled OMOs [right ($R$) and left ($L$)] that are fabricated with slightly different dimensions (i.e. slightly different mechanical frequencies). The optical coupling means the mechanical displacement of one OMO will lead to a force on the other OMO through the optical field. This force is responsible for the effective mechanical coupling between the two OMOs. As the OMOs are pumped by a blue-detuned CW laser into self-sustaining oscillations, the $R$ ($L$) OMO not only experiences the oscillation at its natural frequency but also a modulated optical force at the $L$ ($R$) OMO's mechanical frequency. As the coupling between the two oscillators is increased, each OMO is eventually forced to oscillate at an intermediate frequency between their natural frequencies (${\Omega _R}$ and ${\Omega _L}$), that is, the onset of synchronization \cite{CroZumLif0411,HolMeaMil1105,Hossein08}. We observe both the individual free-running and synchronized oscillation dynamics by switching on and off the purely optical coupling between two OMOs.

Each individual OMO, shown in figure \hyperref[fig:fig1main]{\ref{fig:fig1main}a,b} consists of two suspended vertically stacked Si$_3$N$_4$ disks. As illustrated in figure \hyperref[fig:fig1main]{\ref{fig:fig1main}a}, the optical and mechanical modes of such a cavity are localized around its free-standing edge. The disks are fabricated using standard electron-beam lithography followed by dry and wet etching steps (see Methods). The two disks are $40$~\textmu m in diameter and $210$~nm in thickness, while the air gap between them is $190$~nm wide. Such a small gap and the relative low refractive index of Si$_3$N$_4$ ($n \approx 2.0$) induce a strong optical coupling between the top and bottom disks. The resonant frequency of the optical modes of the stacked disks depend strongly on their separation \cite{WieCheGon0912}; therefore any mechanical vibration that modulates the vertical gap width also modulates the optical resonant frequency; a measure for the efficiency of this process is the optomechanical coupling, defined as ${g_\text{om}} = \partial \omega /\partial x$ where $\omega $ is the optical frequency and $x$ is the mechanical mode amplitude \cite{KipVah07,WieManLee1101,WieCheGon0912}. Our device exhibits a large optomechanical coupling,  calculated to be $g_\text{om}/2\pi =49$~GHz/nm (see SI). The mechanical mode that couples most strongly to the optical field is also illustrated by the deformation of the disks edge in figures \hyperref[fig:fig1main]{\ref{fig:fig1main}a,c} which has a natural frequency of ${\Omega _m}/2\pi  \approx 50.5$~MHz. Note that variations in the fabrication process lead to different mechanical frequencies; indeed we show below that the two cavities are not identical and without the optical coupling they oscillate at different mechanical frequencies.

The two OMOs are separated by a distance of ${d_g}=(400\pm20)$ nm, minimizing direct mechanical coupling. This gap results in evanescent optical coupling between the OMOs when their optical resonant frequencies are close. The optical coupling leads to two optical supermodes spatially spanning both OMOs: a symmetric, lower frequency mode $b_+(t)$ (figure \hyperref[fig:fig1main]{\ref{fig:fig1main}c}) and an anti-symmetric higher frequency mode $b_-(t)$ (figure \hyperref[fig:fig1main]{\ref{fig:fig1main}d}). Their eigenfrequencies are given by $\omega_\pm=\bar{\omega}\pm\kappa/2$, where $\bar{\omega}=(\omega_L+\omega_R)/2$ and $\omega_L$ ($\omega_R$) is the uncoupled optical resonant frequency of the $L$ ($R$) OMO and $\kappa$ is the optical coupling rate: a reflection of the distance between the two cavities. The mechanical modes of each cavity can be approximated by a lumped model consisting of two damped harmonic oscillators, which are driven by the optical supermode forces, 
\begin{equation}
\label{eq:mec_eq1}
\ddot{x}_j+\Gamma_j \dot{x}_j+\Omega_j^2 x_i=F^{(j)}_\text{opt}(x_R,x_L)/m_\text{eff}^{(j)}, \text{for }j,k=L,R
\end{equation}
where $x_j,\Omega_j,\Gamma_j,m_\text{eff}^{(j)}$ represent the mechanical displacement, mechanical resonant frequency, dissipation rate, and effective motional mass of each mechanical degree of freedom. The optical force is proportional to the optical energy stored in the coupled optical modes, which depend both on $x_R$ and $x_L$, i.e. $F^{(j)}_\text{opt}(x_R,x_L)\propto|b_\pm(x_R,x_L)|^2$.  Therefore the optical field not only drives but also mechanically couples each OMO. The nonlinear nature of this driving and coupling force form the basis for the onset of synchronization. In a first order linear approximation when the two OMOs are evenly coupled ($\omega_L=\omega_R$), the effective mechanical coupling force between the two oscillators is given by $F_\text{coup}^{(i)}=-k_{I}x_j + k_Q\dot{x}_j$ where $k_I$ and $k_Q$ are the position and velocity coupling coefficients (See supplementary information [SI] for details). These coupling coefficients are determined by both the input optical power $P_{\textnormal{in}}$ and laser-cavity detuning $\Delta$ as $k_I\propto\frac{P_{\textnormal{in}}\Delta}{((\gamma/2)^2+ \Delta^2)^2}$ and $k_Q\propto\frac{P_{\textnormal{in}}(\gamma/2)\Delta}{((\gamma/2)^2+ \Delta^2)^3}$ in the unresolved side band limit (optical damping rate $\gamma \ll \Omega_{j}$) as in our system. Therefore, by varying $\Delta$ and $P_{\textnormal{in}}$, hence the effective mechanical coupling strength, synchronization of the two OMOs can be captured. 

\begin{figure*}[htbp!]
\begin{center}
\includegraphics[width=0.9\textwidth,angle=0]{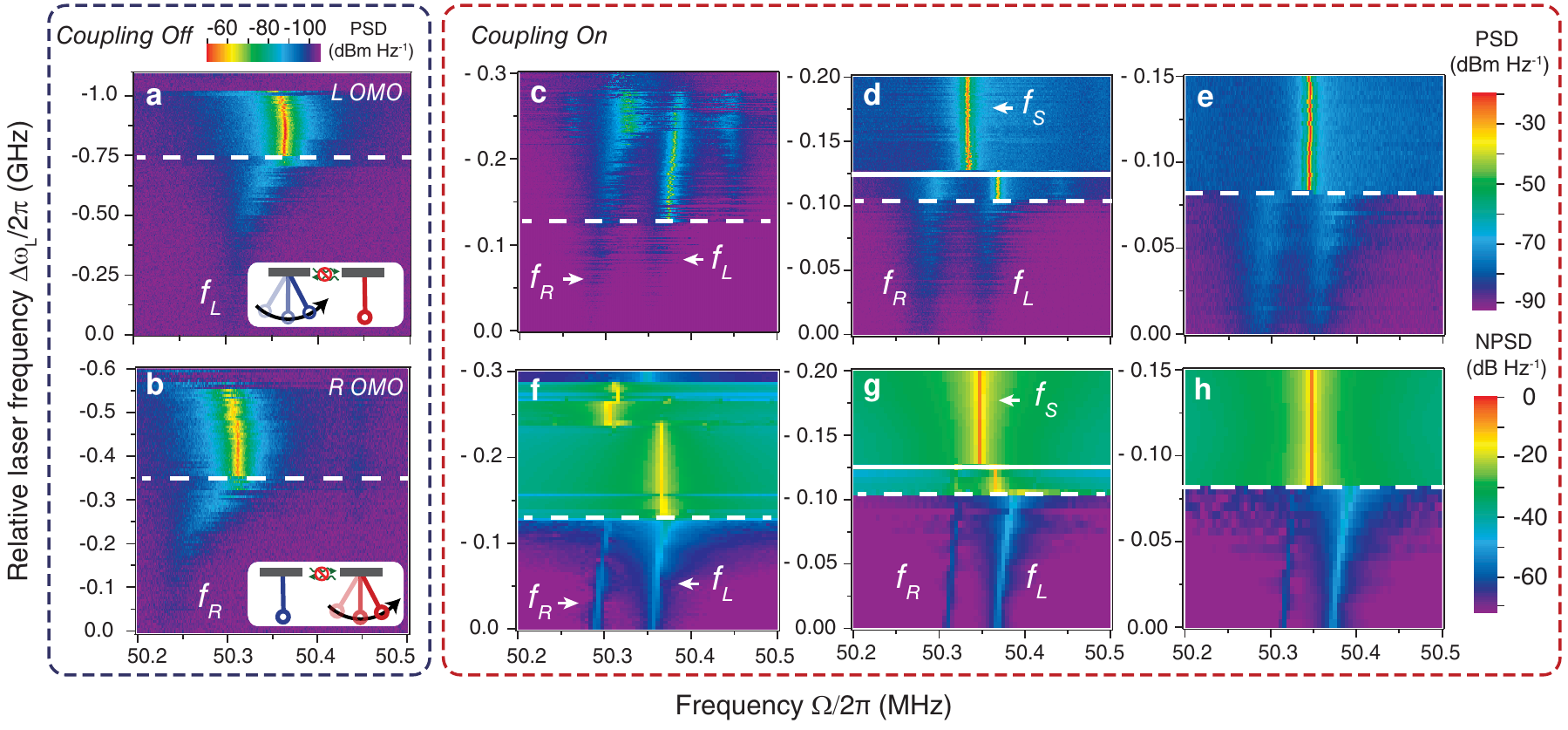}
\caption{\label{fig:fig3main}RF spectra of the OMOs and synchronization (\textbf{a}, \textbf{b}) RF power spectra of cavity $L$ (a) and $R$ (b) as a function of laser frequency when the coupling is turned off.  The horizontal white lines indicate the onset of self-sustaining oscillation.  PSD: power spectral density. (\textbf{c}) When the coupling is turned on, at an input power $P_{in}=(1.8\pm0.2)$~\textmu W cavities $L$ and $R$ do not synchronize and oscillate close to their natural frequencies (see SI). (\textbf{d}) At ${P_{in}}=(11\pm1)$~\textmu W synchronization occurs after the horizontal solid white line. The synchronized frequency appears between the two cavities natural frequencies but only appear after a region of unsynchronized oscillation (between the dashed and solid white lines). (\textbf{e}) The system oscillate directly in a synchronized state at input optical power ${P_{in}}=(14\pm1)$~\textmu W. (\textbf{f},\textbf{g},\textbf{h}) Corresponding numerical simulations for the OMO system based on the lumped harmonic oscillator model illustrated in fig. \hyperref[fig:fig1main]{\ref{fig:fig1main}d}. NPSD: normalized power spectral density.}
\end{center}
\end{figure*}

We experimentally demonstrate that the system can be reconfigured to exhibit either coupled or single OMO dynamics by controlling the spatial distribution of the optical field between the two oscillators. While the distance between the two OMOs is fixed (i.e. fixed $\kappa$), their optical coupling can be turned off (on) through increasing (decreasing) the optical frequency mismatch $\delta=\omega_R-\omega_L$ between them. For large optical frequency mismatch among the two OMOs ($\delta\gg\kappa$) the supermodes reduce to the uncoupled optical modes of the individual OMO, $(b_+,b_- )\rightarrow (a_L,a_R)$. This can be readily seen from the expression of the optical supermodes amplitudes, which are given by linear combinations of the uncoupled modes of the left $a_L(t)$ and right $a_R(t)$ cavities: $b_{\pm}(t) = a_L(t)-a_{R}(t)\mathrm{i}\kappa/(\delta\mp(\delta^2-\kappa^2)^{1/2})$. We tune $\delta$ experimentally using thermo-optic effect, for which the optical frequency dependence on temperature can be approximated as $\omega_j(T_j)=\omega^{(j)}_{0}-g_\text{th} T_j \text{ for } j=L,R$, where $\omega^{(j)}_{0}$ is the intrinsic optical frequency and $g_\text{th}$ is the thermal-optic tuning efficiency. The thermo-optic tuning is accomplished by focusing an out-of-plane laser beam with wavelength 1550 nm on either OMO (figure \hyperref[fig:fig2main]{\ref{fig:fig2main}a}). In order to increase the laser absorption, we deposit a 200 nm layer of chrome in the centre of both OMOs (figure \hyperref[fig:fig2main]{\ref{fig:fig2main}a,b}). As heat is dissipated in the chrome pads, the cavity temperature increases and red shifts the optical resonance of the cavity through thermo-optic effect. A signature that the optical frequencies of both OMOs are matched is given by the almost symmetric resonance dips observed in the optical transmission spectrum (figure \hyperref[fig:fig2main]{\ref{fig:fig2main}b}, \hyperref[fig:fig2main]{\ref{fig:fig2main}d}), which also indicates maximum optical coupling between the cavity optical modes. We show experimentally that the coupling of the optical modes can be continuously tuned through changing the relative cavity temperature as in figure \hyperref[fig:fig2main]{\ref{fig:fig2main}c}. At $\Delta T=0$ we have the maximum optical coupling, whereas for $\Delta T=\pm 25 K$, the relative frequency difference is large ($\delta\gg\kappa$) and the optical mode in eq. \eqref{eq:mec_eq1} does not couple the two OMOs. They follow the usual single-cavity optomechanical dynamics \cite{KipVah07}.

\begin{figure*}[htbp!]
\begin{center}
\includegraphics[width=0.9\textwidth,angle=0]{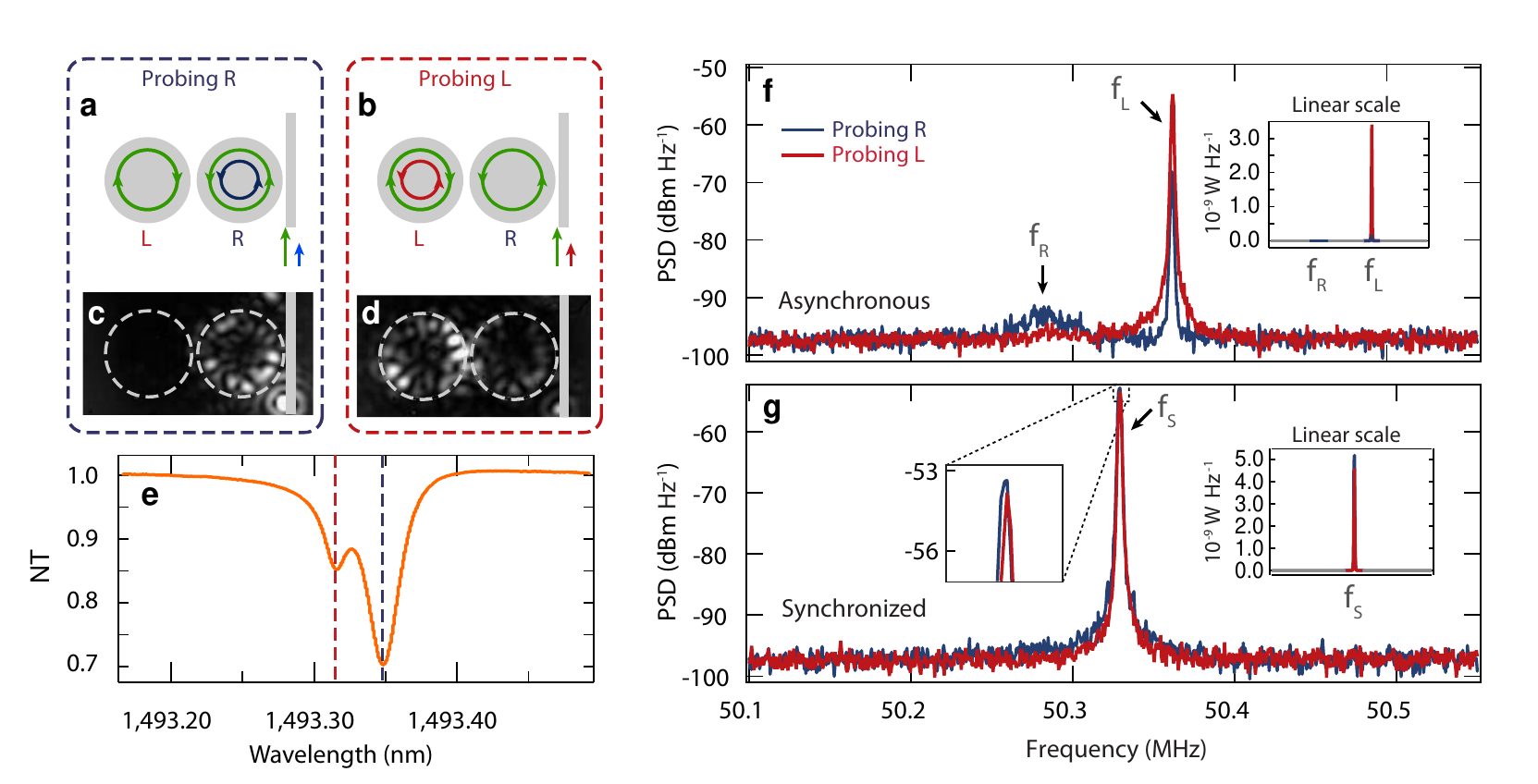}
\caption{Pump-probe measurement of the individual OMOs oscillation when coupled. The input pump power is $P_{in} = (11\pm 1)$~\textmu W as in fig. \hyperref[fig:fig3main]{\ref{fig:fig3main}d}. (\textbf{a},\textbf{b}) Schematic of the pump-probe measurement principle. While the pump laser (green) is symmetrically shared between the two OMOs, the probe laser (blue for probing $R$ and red for $L$) can measure each cavity selectively. (\textbf{c},\textbf{d}) The uneven probe intensity distribution of the cavities, observed by an infrared CCD camera when the pump laser is off.  (\textbf{e}) Normalized transmission (NT) spectrum for the probe resonances, which correspond to the orange resonances shown in fig. \hyperref[fig:fig2main]{\ref{fig:fig2main}b}. The red (blue) dashed line corresponds to the probe wavelength region for probing the $L$ ($R$) OMO, as illustrated in (a,b). (\textbf{f}) The red (blue) curve is the $L$ ($R$) cavity probe transmission RF spectrum, when the pump is in the asynchronous region $- 0.13 < \Delta {\omega _L}/2\pi  <  - 0.10 $ GHz shown in fig. \hyperref[fig:fig3main]{\ref{fig:fig3main}d}; a strong peak at $f_R$ is observed but with very different amplitude for two probing conditions. The right inset figures show the same curves in linear scale, emphasising the large difference between the blue and red curves. (\textbf{g}) Same curves shown in (f) but with the pump laser in the synchronous region $\Delta {\omega _L}/2\pi < -0.13$~GHz of fig. \hyperref[fig:fig3main]{\ref{fig:fig3main}d}. Here both cavities have similar amplitude at $f_S$, which can be clearly noticed in the linear scale inset.}
\label{fig:fig4main}
\end{center}
\end{figure*}

 We first characterize the individual dynamics of the two OMOs by switching their optical coupling off. This is achieved through increasing the heating laser power such that the temperature difference corresponds to the extremities in figure \hyperref[fig:fig2main]{\ref{fig:fig2main}c}. Each cavity is individually excited with a CW laser through a tapered optical fibre. As the laser frequency is tuned (from a higher to a lower frequency) into the optical resonance, the radio-frequency (RF) spectrum of the transmitted laser signal is detected by a photodiode (PD) and recorded using a RF spectrum analyser (RSA). The results revealing the single-cavity optomechanical dynamics are shown in figure \hyperref[fig:fig3main]{\ref{fig:fig3main}a,b}. The mechanical modes have natural mechanical frequencies of $(f_L,f_R)=(\Omega _L,\Omega_R)/2\pi=(50.283,50.219)$~MHz, and intrinsic quality factors of $(Q_{m}^{(L)},Q_{m}^{(R)})=(3.4\pm0.3, 2.3\pm0.2)\times 10^3$. Due to the increased optomechanical back-action and intracavity optical power the OMOs have their frequencies increased (optical spring effect) and amplitudes grown as the laser is tuned into the optical resonance. Above a specific laser-cavity detuning, indicated by the horizontal white dashed lines on figure \hyperref[fig:fig3main]{\ref{fig:fig3main}a,b} the intrinsic mechanical losses are completely suppressed by the optomechanical amplification. At this point the optomechanical resonator starts self-sustaining oscillations and becomes an OMO characterized by sudden linewidth narrowing and oscillation amplitude growth \cite{HeiLudQia1107,MarHarGir0603,HolMeaMil1105}. As the laser frequency sweeps away from the optical resonance it eventually reaches the point of maximum optical power coupled to the cavity; further sweeping can only reduce the optical power inside the cavity and the oscillation vanishes. It is clear from figure \hyperref[fig:fig3main]{\ref{fig:fig3main}a,b} that each cavity has only one mechanical mode in the frequency range of interest. Due to the slight difference in geometry, these frequencies differ by $\Delta f  = f_L - f_R  = (70.0\pm0.5)$~kHz.

We show the onset of spontaneous synchronization by sweeping the CW pump laser across the optical resonance, similarly to the single-cavity measurements above only now the optical coupling is switched on for coupled dynamics. Using the heating laser, we tune the optical coupling to its maximum value, indicated by the dashed-white line ($T_R-T_L=0$) in figure \hyperref[fig:fig2main]{\ref{fig:fig2main}c}. The laser frequency sweeping is performed at various optical power levels corresponding to different effective mechanical coupling strength. The optical power ranges from slightly above the estimated oscillation threshold (i.e weaker mechanical coupling for the $L$ and $R$ OMOs, ${P_{\text{th}}^{(L,R)}} \approx (640,880)$~nW, up to several times their threshold power (i.e. stronger mechanical coupling). At a relative low input power, ${P_\text{in}} = (1.8\pm0.2)~$\textmu W, the mechanical peaks at $f_R$ and $f_L$ are simultaneously observed on the RF spectrum shown in figure \hyperref[fig:fig3main]{\ref{fig:fig3main}c}, below the dashed-white line. When the laser frequency is closer to the optical resonant frequency, more energy is available and the $L$ OMO starts self-sustaining oscillation. Since cavity $R$ has a higher oscillation threshold, due to its lower mechanical quality factor, it requires more optical power and only oscillates at a redder detuning; it can be noticed from figure \hyperref[fig:fig3main]{\ref{fig:fig3main}c} that both OMOs oscillate close to their natural frequency. Therefore they exhibit asynchronous oscillations at this lowest power level. At a higher input optical power level of ${P_\text{in}} = (11\pm1)$~\textmu W, the first oscillation takes place at $\Delta {\omega _L}/2\pi  \approx - 0.10$~GHz, and similarly to the case shown in figure \hyperref[fig:fig3main]{\ref{fig:fig3main}c}, the $L$ OMO oscillates first. However, as the laser frequency further moves into the optical resonance, there is enough energy for both OMOs to start self-sustaining oscillations; the two OMOs spontaneously oscillate in unison at an intermediate frequency of $f_S={\Omega _S}/2\pi  = 50.37$~MHz due to the increased effective mechanical coupling, which is a clear sign of synchronization. At this time, the output optical RF power is increase by more than 5 dB in comparison with the $L$ OMO oscillating only case showing that the two OMOs are phase-locked. At an even higher optical input power, ${P_\text{in}} = (14\pm1)$~\textmu W, the OMOs do not oscillate individually, instead they go directly into synchronized oscillations above the white-dashed line in figure \hyperref[fig:fig3main]{\ref{fig:fig3main}e}. We confirm that the OMOs are indeed synchronized by performing numerical simulations corresponding to each of the power levels we tested.  The simulated spectra in figure \hyperref[fig:fig3main]{\ref{fig:fig3main}f,g,h} exhibit all the essential features observed and show good agreement with the measured spectra. It also allows us to confirm under which conditions the two OMOs are indeed oscillating (see Methods and SI). 

To experimentally verify that both structures are indeed oscillating at the synchronized frequency, we probe the mechanical oscillation of each cavity individually. This demonstrates that the single oscillation peaks observed in figures \hyperref[fig:fig3main]{\ref{fig:fig3main}d,e} are not caused by one OMO resonantly driving the other; it also verifies that amplitude death of one of the OMOs does not occur, a known phenomenon in coupled nonlinear oscillators \cite{MirStr9007}. The transmitted pump laser signal only provides only global information of the coupled OMO system, it does not distinguishes the individual contribution from each OMO to the synchronized signal. To overcome this we used a weak CW probe laser, as shown in the setup in figure \hyperref[fig:fig2main]{\ref{fig:fig2main}a} to excite an optical resonant mode that is not strongly coupled between the two OMOs; this scheme illustrated in figure \hyperref[fig:fig4main]{\ref{fig:fig4main}a,b} allows us to selectively probe the oscillations of the $L$ or $R$ OMO. Figure \hyperref[fig:fig4main]{\ref{fig:fig4main}c,d} show the uneven light intensity distribution that can be directly observed by capturing the scattered light with an infrared camera. The asymmetric splitting is also evident on the probe transmission shown in figure \hyperref[fig:fig4main]{\ref{fig:fig4main}e}. While these probe optical modes exhibit a low optical quality factor ($Q_{opt} \approx 4 \times {10^4}$) that minimizes probe-induced perturbations to the mechanical oscillations, the pump laser power and sweep is identical to the one used figure \hyperref[fig:fig3main]{\ref{fig:fig3main}d}. When the $L$ OMO is probed, and the pump detuning range is between the dashed and solid lines in figure \hyperref[fig:fig3main]{\ref{fig:fig3main}d}, the probe RF spectrum shows a strong peak at ${f_L}$, which is shown in the red curve in figure \hyperref[fig:fig4main]{\ref{fig:fig4main}f}. When the $R$ OMO is probed, a peak also appears at this frequency, but it is 13 dB weaker as shown in the blue curve in figure \hyperref[fig:fig4main]{\ref{fig:fig4main}f}; a weak peak at ${f_R}$ can also be noticed on the blue curve, indicating small amplitude oscillatios of the $R$ OMO.  These results confirm that the oscillation state is very asynchronous in this detuning range with the $L$ OMO oscillating at much larger amplitude. When the pump laser detuning is above the horizontal solid line in figure \hyperref[fig:fig3main]{\ref{fig:fig3main}d} there is only a single RF peak at the synchronized frequency ${f_S}$ when probing either OMO (figure \hyperref[fig:fig4main]{\ref{fig:fig4main}g}); moreover, they differ in amplitude by less than 0.5 dB. This shows that both cavities are indeed oscillating with similar strength at the synchronized frequency.

We have demonstrated the onset of synchronization between two optomechanical oscillators coupled only through the optical radiation field. The ability to control the coupling strength are promising for realizing oscillator networks in which the oscillators can be addressed individually. Furthermore, established and future micro-photonics techniques such as electro-optic and thermo-optic techniques can now be extended to switch, filter and phase shift the coupling of these oscillators. Here we demonstrated coupling the near field between oscillators which can be switched on and off by thermo-optical means. In order to achieve long range coupling of mechanical oscillators, optical waveguides and optical fibres could be used enabling oscillator networks spread over large areas only limited by optical waveguide/fibre losses. Optically mediated mechanical coupling will also remove the restrictions of neighbourhood while creating 1D/2D/3D mechanical oscillator arrays \cite{ArlMyeRou1104}. Using long range, directional and controllable mechanical coupling, synchronized optomechanical systems may enable a new class of devices in sensing, signal processing and on-chip non-linear dynamical systems \cite{Mar1110}.

\tocless{\section*{Methods}}

\textbf{Coupled optical cavities}
A detailed model of this coupled optomechanical system is discussed in the SI; here we describe a simplified version that capture our system's essential aspects. The optical mode amplitudes are approximated by two coupled harmonic oscillator equations given by \cite{Hau84},
\begin{equation}
\label{eq:opt_eq1}
\begin{array}{lcl}
\dot{a}_j=\mathrm{i}\Delta_j a_j - (\gamma/2)a_j+\mathrm{i}\kappa a_k+
\sqrt{\gamma_e}s_j,\vspace{6pt} \\
\hspace{1.3in} \text{for }j,k=L,R, j\neq k\\
\end{array}
\end{equation}
where $\Delta_{j}=\omega-\omega_{j}(x_{j},T_{j})$ is the laser-cavity frequency detuning, $\omega$ is the optical frequency of the CW driving laser, $\omega_{j}(x_{j},T_{j})$ is the optical frequency of the each uncoupled optical mode ($a_{R,L}$), which depends both on each cavity temperature ($T_{R,L}$) and mechanical mode amplitude ($x_{R,L})$. The optical dissipation rate is given by $\gamma$ and is assumed to be the same for both cavities; the coupling rate $\kappa$ measures the evanescent field interaction of the two modes and couples the two optical cavities. The CW laser drive $\sqrt{\gamma_e}s_j$, which depends on the cavity coupling to the tapered fibre $\gamma_e$, only drives the $a_R$ mode ($\sqrt{\gamma_e}s_L=0$), resembling the experimental configuration shown in figure \hyperref[fig:fig2main]{\ref{fig:fig2main}a} where the tapered fibre excites only the rightmost cavity. Eqs. \eqref{eq:opt_eq1} can be diagonalized to yield the coupled optical modes (supermodes) we described in the main text.\\
\textbf{Device Fabrication} The two $210$~nm thick stoichiometric Si$_3$N$_4$ films are deposited using low-pressure chemical vapour deposition (LPCVD). The $190$~nm SiO$_2$ layer is deposited by plasma-enhanced chemical vapour deposition (PECVD). The underlying substrate is a $4$ \textmu m SiO$_2$ formed by thermal oxidation of a silicon wafer. The OMOs are defined by electron beam lithography which is then patterned by reactive ion etching. The heater pads are subsequently defined by photolithography lift-off process. After defining the circular pads with lift-off resist, $200$ nm of chrome is deposited on the device using electron beam evaporation and the residual chrome is lift-off afterwards. In order to release the structure, the device is immersed in buffered hydrofluoric acid ($6:1$) for an isotropic etch of the SiO$_2$ in between the disks and the substrate layer. The device is then dried with a critical point dryer to avoid stiction between the two Si$_3$N$_4$ disks.\\
\textbf{Experimental setup} The schematic for testing the OMO system is illustrated in figure \hyperref[fig:fig2main]{\ref{fig:fig2main}a}. Two tunable external cavity diode lasers are combined using a 3dB directional coupler to an optical fiber that is fed into a vacuum probe station. Inside the vacuum chamber, the tapered fiber is positioned close to the OMO of interest to allow evanecent coupling using a micropositioning system. The output light is then splitted by a WDM splitter to a New Focus 1811 (125 MHz bandwidth) photodetector. Since the power level we use to test for our device is low, an erbium doped fibre preamplifer is used to amplify the output signal and improve signal-to-noise ratio in the detector. The electronic signal from the detector is split and fed to an oscilloscope to observe the time waveform and to RSA for the frequency spectrum. To obtain the RF map, the laser is configured to sweep from the blue side of the resonance to the red side in a stepwise fashion by applying an external voltage to the laser cavity piezo-transducer. At each frequency step, a snap-shot of the RF spectrum is recorded with 1 kHz resolution bandwidth and 100 Hz video bandwidth.\\
\textbf{Simulation} The optical and mechanical quality factors are obtained by non-linear least square fitting to the measured optical and RF spectrum. The measured mechanical and optical frequencies, and their respective quality factors are fed as parameters to the lumped model described by eqs. \eqref{eq:mec_eq1} and \eqref{eq:opt_eq1}. Other parameters such as the effective motional mass $m_\text{eff}$ and the optomechanical coupling $g_\text{om}$  are obtained from finite element simulations (FEM). From these measured and calculated parameters we estimate the threshold power from the expression $P_\text{th} =m_\text{eff}\Omega_m \omega_{0}^{4}(4 Q_{m} Q^{3}(g_\text{om})^2\eta_c )^{-1} $, where $\eta_c=\gamma_e/(\gamma)$ is the coupling ideality factor. Due to nonlinear dependence on most parameters, such as $Q$ and $g_\text{om}$ the error propagation is large in the estimated threshold power. The error propagation in the threshold estimation is detailed in the SI, where we estimate an overall error of $\delta P_\text{th}/P_\text{th}=35\%$. To obtain the density plots shown in figure \hyperref[fig:fig3main]{\ref{fig:fig3main}} we feed the dynamical model described in the main text, and detailed in the SI, with the measured and calculated parameters. The simulations are performed at various power levels close to the experimentally measured power. The optimum power matching was chosen to match the laser frequencies at which the bifurcations take place in the experimental data.

\tocless{\section*{Acknowledgement}}

This work was supported in part by the National Science Foundation under grant 0928552. The authors gratefully acknowledge the partial support from Cornell Center for Nanoscale Systems which is funded by the National Science Foundation and funding from IGERT: A Graduate Traineeship in Nanoscale Control of Surfaces and Interfaces (DEG-0654193). This work was performed in part at the Cornell Nano-Scale Science \& Technology Facility (a member of the National Nanofabrication Users Network) which is supported by National Science Foundation, its users, Cornell University and Industrial users. G.S.W acknowledges FAPESP and CNPq INCT Fotonicom for financial support in Brazil.
We acknowledge Paulo Nussenzveig, Richard Rand and Steven Strogatz for fruitful discussion about our results.

\tocless{\section*{Author Contributions}}

M. Z. and G.W designed, fabricated and tested the devices. S. M. designed and simulated the devices. A. B. designed the experimental setup and helped testing. P.M. and M.L supervised all stages of the experiment. All authors discussed the results and their implications and contributed to writing this manuscript.

\tocless{\section*{Competing Interest}}
%\begin{center}
%\end{center}
The authors declare that they have no competing financial interests. 
Correspondence and requests for materials should be addressed to M.L. (email: ml292@cornell.edu)

\tocless{\bibliography{synchronization}}

\begin{thebibliography}{10}
\expandafter\ifx\csname url\endcsname\relax
  \def\url#1{\texttt{#1}}\fi
\expandafter\ifx\csname urlprefix\endcsname\relax\def\urlprefix{URL }\fi
\providecommand{\bibinfo}[2]{#2}
\providecommand{\eprint}[2][]{\url{#2}}

\bibitem{BucBuc6803}
\bibinfo{author}{Buck, J.} \& \bibinfo{author}{Buck, E.}
\newblock \bibinfo{title}{Mechanism of rhythmic synchronous flashing of
  fireflies. fireflies of southeast asia may use anticipatory time-measuring in
  synchronizing their flashing}.
\newblock \emph{\bibinfo{journal}{Science}} \textbf{\bibinfo{volume}{159}},
  \bibinfo{pages}{1319--27} (\bibinfo{year}{1968}).

\bibitem{Pes75}
\bibinfo{author}{Peskin, C.~S.}
\newblock \emph{\bibinfo{title}{Mathematical aspects of heart physiology}}
  (\bibinfo{publisher}{Courant Institute of Mathematical Sciences, New York
  University}, \bibinfo{address}{New York}, \bibinfo{year}{1975}).

\bibitem{WhiTraJef9502}
\bibinfo{author}{Whittington, M.~A.}, \bibinfo{author}{Traub, R.~D.} \&
  \bibinfo{author}{Jefferys, J.~G.}
\newblock \bibinfo{title}{Synchronized oscillations in interneuron networks
  driven by metabotropic glutamate receptor activation}.
\newblock \emph{\bibinfo{journal}{Nature}} \textbf{\bibinfo{volume}{373}},
  \bibinfo{pages}{612--5} (\bibinfo{year}{1995}).

\bibitem{Str03}
\bibinfo{author}{Strogatz, S.~H.}
\newblock \emph{\bibinfo{title}{Sync: the emerging science of spontaneous
  order}} (\bibinfo{publisher}{Hyperion}, \bibinfo{address}{New York},
  \bibinfo{year}{2003}), \bibinfo{edition}{1st ed} edn.

\bibitem{Bah09}
\bibinfo{author}{Bahder, T.~B.}
\newblock \emph{\bibinfo{title}{Clock synchronization and navigation in the
  vicinity of the earth}} (\bibinfo{publisher}{Nova Science},
  \bibinfo{address}{New York}, \bibinfo{year}{2009}).

\bibitem{Bre02}
\bibinfo{author}{Bregni, S.}
\newblock \emph{\bibinfo{title}{Synchronization of digital telecommunications
  networks}} (\bibinfo{publisher}{Wiley}, \bibinfo{address}{Chichester},
  \bibinfo{year}{2002}).

\bibitem{KakPufRip0509}
\bibinfo{author}{Kaka, S.} \emph{et~al.}
\newblock \bibinfo{title}{Mutual phase-locking of microwave spin torque
  nano-oscillators}.
\newblock \emph{\bibinfo{journal}{Nature}} \textbf{\bibinfo{volume}{437}},
  \bibinfo{pages}{389--92} (\bibinfo{year}{2005}).

\bibitem{MahYam0805}
\bibinfo{author}{Mahboob, I.} \& \bibinfo{author}{Yamaguchi, H.}
\newblock \bibinfo{title}{Bit storage and bit flip operations in an
  electromechanical oscillator}.
\newblock \emph{\bibinfo{journal}{Nature Nanotechnology}}
  \textbf{\bibinfo{volume}{3}}, \bibinfo{pages}{275--279}
  (\bibinfo{year}{2008}).

\bibitem{BagPooLi1110}
\bibinfo{author}{Bagheri, M.}, \bibinfo{author}{Poot, M.}, \bibinfo{author}{Li,
  M.}, \bibinfo{author}{Pernice, W. P.~H.} \& \bibinfo{author}{Tang, H.~X.}
\newblock \bibinfo{title}{Dynamic manipulation of nanomechanical resonators in
  the high-amplitude regime and non-volatile mechanical memory operation}.
\newblock \emph{\bibinfo{journal}{Nature Nanotechnology}}
  (\bibinfo{year}{2011}).

\bibitem{Str98}
\bibinfo{author}{Watts, D.~J.} \& \bibinfo{author}{Strogatz, S.~H.}
\newblock \bibinfo{title}{Collective dynamics of small-world networks}.
\newblock \emph{\bibinfo{journal}{Nature}} \textbf{\bibinfo{volume}{393}},
  \bibinfo{pages}{440--442} (\bibinfo{year}{1998}).
\newblock \bibinfo{note}{10.1038/30918}.

\bibitem{HopIzh0102}
\bibinfo{author}{Hoppensteadt, F.} \& \bibinfo{author}{Izhikevich, E.}
\newblock \bibinfo{title}{Synchronization of mems resonators and mechanical
  neurocomputing}.
\newblock \emph{\bibinfo{journal}{Ieee Transactions On Circuits and Systems
  I-Fundamental Theory and Applications}} \textbf{\bibinfo{volume}{48}},
  \bibinfo{pages}{133--138} (\bibinfo{year}{2001}).

\bibitem{ManRizEng0509}
\bibinfo{author}{Mancoff, F.}, \bibinfo{author}{Rizzo, N.},
  \bibinfo{author}{Engel, B.} \& \bibinfo{author}{Tehrani, S.}
\newblock \bibinfo{title}{Phase-locking in double-point-contact spin-transfer
  devices}.
\newblock \emph{\bibinfo{journal}{Nature}} \textbf{\bibinfo{volume}{437}},
  \bibinfo{pages}{393--395} (\bibinfo{year}{2005}).

\bibitem{ShiImbMoh0704}
\bibinfo{author}{Shim, S.-B.}, \bibinfo{author}{Imboden, M.} \&
  \bibinfo{author}{Mohanty, P.}
\newblock \bibinfo{title}{Synchronized oscillation in coupled nanomechanical
  oscillators}.
\newblock \emph{\bibinfo{journal}{Science}} \textbf{\bibinfo{volume}{316}},
  \bibinfo{pages}{95--99} (\bibinfo{year}{2007}).

\bibitem{ZalAubPan0310}
\bibinfo{author}{Zalalutdinov, M.} \emph{et~al.}
\newblock \bibinfo{title}{Frequency entrainment for micromechanical
  oscillator}.
\newblock \emph{\bibinfo{journal}{Applied Physics Letters}}
  \textbf{\bibinfo{volume}{83}}, \bibinfo{pages}{3281--3283}
  (\bibinfo{year}{2003}).

\bibitem{HeiLudQia1107}
\bibinfo{author}{Heinrich, G.}, \bibinfo{author}{Ludwig, M.},
  \bibinfo{author}{Qian, J.}, \bibinfo{author}{Kubala, B.} \&
  \bibinfo{author}{Marquardt, F.}
\newblock \bibinfo{title}{Collective dynamics in optomechanical arrays}.
\newblock \emph{\bibinfo{journal}{Physical Review Letters}}
  \textbf{\bibinfo{volume}{107}}, \bibinfo{pages}{043603}
  (\bibinfo{year}{2011}).

\bibitem{MarHarGir0603}
\bibinfo{author}{Marquardt, F.}, \bibinfo{author}{Harris, J.} \&
  \bibinfo{author}{Girvin, S.}
\newblock \bibinfo{title}{Dynamical multistability induced by radiation
  pressure in high-finesse micromechanical optical cavities}.
\newblock \emph{\bibinfo{journal}{Physical Review Letters}}
  \textbf{\bibinfo{volume}{96}}, \bibinfo{pages}{103901}
  (\bibinfo{year}{2006}).

\bibitem{Mar1110}
\bibinfo{author}{Marquardt, F.}
\newblock \bibinfo{title}{Quantum mechanics: The gentle cooling touch of
  light}.
\newblock \emph{\bibinfo{journal}{Nature}} \textbf{\bibinfo{volume}{478}},
  \bibinfo{pages}{47--8} (\bibinfo{year}{2011}).

\bibitem{BukRou0212}
\bibinfo{author}{Buks, E.} \& \bibinfo{author}{Roukes, M.}
\newblock \bibinfo{title}{Electrically tunable collective response in a coupled
  micromechanical array}.
\newblock \emph{\bibinfo{journal}{Journal of Microelectromechanical Systems}}
  \textbf{\bibinfo{volume}{11}}, \bibinfo{pages}{802--807}
  (\bibinfo{year}{2002}).

\bibitem{KarCroRou0904}
\bibinfo{author}{Karabalin, R.~B.}, \bibinfo{author}{Cross, M.~C.} \&
  \bibinfo{author}{Roukes, M.~L.}
\newblock \bibinfo{title}{Nonlinear dynamics and chaos in two coupled
  nanomechanical resonators}.
\newblock \emph{\bibinfo{journal}{Physical Review B}}
  \textbf{\bibinfo{volume}{79}}, \bibinfo{pages}{165309}
  (\bibinfo{year}{2009}).

\bibitem{KipVah07}
\bibinfo{author}{Kippenberg, T.} \& \bibinfo{author}{Vahala, K.}
\newblock \bibinfo{title}{Cavity opto-mechanics}.
\newblock \emph{\bibinfo{journal}{Optics Express}}
  \textbf{\bibinfo{volume}{15}}, \bibinfo{pages}{17172--17205}
  (\bibinfo{year}{2007}).

\bibitem{BraStrVya0109}
\bibinfo{author}{Braginsky, V.}, \bibinfo{author}{Strigin, S.} \&
  \bibinfo{author}{Vyatchanin, S.}
\newblock \bibinfo{title}{Parametric oscillatory instability in fabry-perot
  interferometer}.
\newblock \emph{\bibinfo{journal}{Physics Letters a}}
  \textbf{\bibinfo{volume}{287}}, \bibinfo{pages}{331--338}
  (\bibinfo{year}{2001}).

\bibitem{ArcCohBri06}
\bibinfo{author}{Arcizet, O.}, \bibinfo{author}{Cohadon, P.},
  \bibinfo{author}{Briant, T.}, \bibinfo{author}{Pinard, M.} \&
  \bibinfo{author}{Heidmann}.
\newblock \bibinfo{title}{Radiation-pressure cooling and optomechanical
  instability of a micromirror}.
\newblock \emph{\bibinfo{journal}{Nature}} \textbf{\bibinfo{volume}{444}},
  \bibinfo{pages}{71--74} (\bibinfo{year}{2006}).

\bibitem{SheGraMow0405}
\bibinfo{author}{Sheard, B.}, \bibinfo{author}{Gray, M.},
  \bibinfo{author}{Mow-Lowry, C.}, \bibinfo{author}{McClelland, D.} \&
  \bibinfo{author}{Whitcomb, S.}
\newblock \bibinfo{title}{Observation and characterization of an optical
  spring}.
\newblock \emph{\bibinfo{journal}{Physical Review a}}
  \textbf{\bibinfo{volume}{69}}, \bibinfo{pages}{051801}
  (\bibinfo{year}{2004}).

\bibitem{CarRokYan05}
\bibinfo{author}{Carmon, T.}, \bibinfo{author}{Rokhsari, H.},
  \bibinfo{author}{Yang, L.}, \bibinfo{author}{Kippenberg, T.~J.} \&
  \bibinfo{author}{Vahala, K.~J.}
\newblock \bibinfo{title}{Temporal behavior of radiation-pressure-induced
  vibrations of an optical microcavity phonon mode}.
\newblock \emph{\bibinfo{journal}{Physical Review Letters}}
  \textbf{\bibinfo{volume}{94}}, \bibinfo{pages}{223902}
  (\bibinfo{year}{2005}).

\bibitem{GruLeePai1002}
\bibinfo{author}{Grudinin, I.~S.}, \bibinfo{author}{Lee, H.},
  \bibinfo{author}{Painter, O.} \& \bibinfo{author}{Vahala, K.~J.}
\newblock \bibinfo{title}{Phonon laser action in a tunable two-level system}.
\newblock \emph{\bibinfo{journal}{Physical Review Letters}}
  \textbf{\bibinfo{volume}{104}}, \bibinfo{pages}{083901}
  (\bibinfo{year}{2010}).

\bibitem{ManWeiLip11}
\bibinfo{author}{Manipatruni, S.}, \bibinfo{author}{Wiederhecker, G.} \&
  \bibinfo{author}{Lipson, M.}
\newblock \bibinfo{title}{Long-range synchronization of optomechanical
  structures}.
\newblock In \emph{\bibinfo{booktitle}{Quantum Electronics and Laser Science
  Conference}}, \bibinfo{pages}{QWI1} (\bibinfo{publisher}{Optical Society of
  America}, \bibinfo{year}{2011}).

\bibitem{HolMeaMil1105}
\bibinfo{author}{{Holmes}, C.~A.}, \bibinfo{author}{{Meaney}, C.~P.} \&
  \bibinfo{author}{{Milburn}, G.~J.}
\newblock \bibinfo{title}{{Multi-stability and synchronization of many
  nano-mechanical resonators coupled via a cavity field}}.
\newblock \emph{\bibinfo{journal}{arXiv e-prints: 1105.2086}}
  (\bibinfo{year}{2011}).

\bibitem{CroZumLif0411}
\bibinfo{author}{Cross, M.}, \bibinfo{author}{Zumdieck, A.},
  \bibinfo{author}{Lifshitz, R.} \& \bibinfo{author}{Rogers, J.}
\newblock \bibinfo{title}{Synchronization by nonlinear frequency pulling}.
\newblock \emph{\bibinfo{journal}{Physical Review Letters}}
  \textbf{\bibinfo{volume}{93}}, \bibinfo{pages}{224101}
  (\bibinfo{year}{2004}).

\bibitem{Hossein08}
\bibinfo{author}{Hossein-Zadeh, M.} \& \bibinfo{author}{Vahala, K.}
\newblock \bibinfo{title}{Observation of injection locking in an optomechanical
  rf oscillator}.
\newblock \emph{\bibinfo{journal}{Applied Physics Letters}}
  \textbf{\bibinfo{volume}{93}}, \bibinfo{pages}{191115--191115--3}
  (\bibinfo{year}{2008}).

\bibitem{WieCheGon0912}
\bibinfo{author}{Wiederhecker, G.~S.}, \bibinfo{author}{Chen, L.},
  \bibinfo{author}{Gondarenko, A.} \& \bibinfo{author}{Lipson, M.}
\newblock \bibinfo{title}{Controlling photonic structures using optical
  forces}.
\newblock \emph{\bibinfo{journal}{Nature}} \textbf{\bibinfo{volume}{462}},
  \bibinfo{pages}{633--U103} (\bibinfo{year}{2009}).

\bibitem{WieManLee1101}
\bibinfo{author}{Wiederhecker, G.~S.}, \bibinfo{author}{Manipatruni, S.},
  \bibinfo{author}{Lee, S.} \& \bibinfo{author}{Lipson, M.}
\newblock \bibinfo{title}{Broadband tuning of optomechanical cavities}.
\newblock \emph{\bibinfo{journal}{Optics Express}}
  \textbf{\bibinfo{volume}{19}}, \bibinfo{pages}{2782--2790}
  (\bibinfo{year}{2011}).

\bibitem{MirStr9007}
\bibinfo{author}{Mirollo, R.} \& \bibinfo{author}{Strogatz, S.}
\newblock \bibinfo{title}{Amplitude death in an array of limit-cycle
  oscillators}.
\newblock \emph{\bibinfo{journal}{Journal of Statistical Physics}}
  \textbf{\bibinfo{volume}{60}}, \bibinfo{pages}{245--262}
  (\bibinfo{year}{1990}).

\bibitem{ArlMyeRou1104}
\bibinfo{author}{Arlett, J.~L.}, \bibinfo{author}{Myers, E.~B.} \&
  \bibinfo{author}{Roukes, M.~L.}
\newblock \bibinfo{title}{Comparative advantages of mechanical biosensors}.
\newblock \emph{\bibinfo{journal}{Nature Nanotechnology}}
  \textbf{\bibinfo{volume}{6}}, \bibinfo{pages}{203--15}
  (\bibinfo{year}{2011}).

\bibitem{Hau84}
\bibinfo{author}{Haus, H.~A.}
\newblock \emph{\bibinfo{title}{Waves and fields in optoelectronics}}
  (\bibinfo{publisher}{Prentice-Hall}, \bibinfo{address}{Englewood Cliffs, NJ},
  \bibinfo{year}{1984}).

\bibitem{AneArcUnt0912}
\bibinfo{author}{Anetsberger, G.} \emph{et~al.}
\newblock \bibinfo{title}{Near-field cavity optomechanics with nanomechanical
  oscillators}.
\newblock \emph{\bibinfo{journal}{Nature Physics}}
  \textbf{\bibinfo{volume}{5}}, \bibinfo{pages}{909--914}
  (\bibinfo{year}{2009}).

\bibitem{EicChaSaf0910}
\bibinfo{author}{Eichenfield, M.}, \bibinfo{author}{Chan, J.},
  \bibinfo{author}{Safavi-Naeini, A.~H.}, \bibinfo{author}{Vahala, K.~J.} \&
  \bibinfo{author}{Painter, O.}
\newblock \bibinfo{title}{Modeling dispersive coupling and losses of localized
  optical and mechanical modes in optomechanical crystals}.
\newblock \emph{\bibinfo{journal}{Optics Express}}
  \textbf{\bibinfo{volume}{17}}, \bibinfo{pages}{20078--20098}
  (\bibinfo{year}{2009}).

\bibitem{JohIbaSko0206}
\bibinfo{author}{Johnson, S.} \emph{et~al.}
\newblock \bibinfo{title}{Perturbation theory for maxwell's equations with
  shifting material boundaries}.
\newblock \emph{\bibinfo{journal}{Physical Review E}}
  \textbf{\bibinfo{volume}{65}}, \bibinfo{pages}{066611}
  (\bibinfo{year}{2002}).

\bibitem{GorPryIlc0006}
\bibinfo{author}{Gorodetsky, M.}, \bibinfo{author}{Pryamikov, A.} \&
  \bibinfo{author}{Ilchenko, V.}
\newblock \bibinfo{title}{Rayleigh scattering in high-q microspheres}.
\newblock \emph{\bibinfo{journal}{Journal of the Optical Society of America
  B-Optical Physics}} \textbf{\bibinfo{volume}{17}},
  \bibinfo{pages}{1051--1057} (\bibinfo{year}{2000}).

\bibitem{SpiKipPai0307}
\bibinfo{author}{Spillane, S.}, \bibinfo{author}{Kippenberg, T.},
  \bibinfo{author}{Painter, O.} \& \bibinfo{author}{Vahala, K.}
\newblock \bibinfo{title}{Ideality in a fiber-taper-coupled microresonator
  system for application to cavity quantum electrodynamics}.
\newblock \emph{\bibinfo{journal}{Physical Review Letters}}
  \textbf{\bibinfo{volume}{91}}, \bibinfo{pages}{043902}
  (\bibinfo{year}{2003}).

\bibitem{Law9503}
\bibinfo{author}{Law, C.~K.}
\newblock \bibinfo{title}{Interaction between a moving mirror and radiation
  pressure: A hamiltonian formulation}.
\newblock \emph{\bibinfo{journal}{Phys. Rev. A}} \textbf{\bibinfo{volume}{51}},
  \bibinfo{pages}{2537--2541} (\bibinfo{year}{1995}).

\bibitem{EicCamCha0905}
\bibinfo{author}{Eichenfield, M.}, \bibinfo{author}{Camacho, R.},
  \bibinfo{author}{Chan, J.}, \bibinfo{author}{Vahala, K.~J.} \&
  \bibinfo{author}{Painter, O.}
\newblock \bibinfo{title}{A picogram- and nanometre-scale photonic-crystal
  optomechanical cavity}.
\newblock \emph{\bibinfo{journal}{Nature}} \textbf{\bibinfo{volume}{459}},
  \bibinfo{pages}{550--555} (\bibinfo{year}{2009}).
\newblock \urlprefix\url{http://dx.doi.org/10.1038/nature08061}.

\bibitem{SAU9010}
\bibinfo{author}{Saulson, P.}
\newblock \bibinfo{title}{Thermal noise in mechanical experiments}.
\newblock \emph{\bibinfo{journal}{Physical Review D}}
  \textbf{\bibinfo{volume}{42}}, \bibinfo{pages}{2437--2445}
  (\bibinfo{year}{1990}).

\bibitem{HauJanBal0611}
\bibinfo{author}{Hauschildt, B.}, \bibinfo{author}{Janson, N.~B.},
  \bibinfo{author}{Balanov, A.} \& \bibinfo{author}{Schoell, E.}
\newblock \bibinfo{title}{Noise-induced cooperative dynamics and its control in
  coupled neuron models}.
\newblock \emph{\bibinfo{journal}{Physical Review E}}
  \textbf{\bibinfo{volume}{74}}, \bibinfo{pages}{051906}
  (\bibinfo{year}{2006}).

\bibitem{HoeSuaSan0109}
\bibinfo{author}{Hoeye, S.}, \bibinfo{author}{Suarez, A.} \&
  \bibinfo{author}{Sancho, S.}
\newblock \bibinfo{title}{Analysis of noise effects on the nonlinear dynamics
  of synchronized oscillators}.
\newblock \emph{\bibinfo{journal}{Ieee Microwave and Wireless Components
  Letters}} \textbf{\bibinfo{volume}{11}}, \bibinfo{pages}{376--378}
  (\bibinfo{year}{2001}).

\bibitem{BurBurHig0612}
\bibinfo{author}{Burrage, K.}, \bibinfo{author}{Burrage, P.},
  \bibinfo{author}{Higham, D.~J.}, \bibinfo{author}{Kloeden, P.~E.} \&
  \bibinfo{author}{Platen, E.}
\newblock \bibinfo{title}{Comment on "numerical methods for stochastic
  differential equations"}.
\newblock \emph{\bibinfo{journal}{Physical Review E}}
  \textbf{\bibinfo{volume}{74}}, \bibinfo{pages}{068701}
  (\bibinfo{year}{2006}).

\bibitem{KloPla99}
\bibinfo{author}{Kloeden, P.~E.} \& \bibinfo{author}{Platen, E.}
\newblock \emph{\bibinfo{title}{Numerical solution of stochastic differential
  equations}}, vol.~\bibinfo{volume}{23} (\bibinfo{publisher}{Springer},
  \bibinfo{address}{Berlin}, \bibinfo{year}{1999}), \bibinfo{edition}{corr. 3rd
  print} edn.
\newblock
  \urlprefix\url{http://www.loc.gov/catdir/enhancements/fy0812/00266312-d.html}.

\bibitem{Wil0407}
\bibinfo{author}{Wilkie, J.}
\newblock \bibinfo{title}{Numerical methods for stochastic differential
  equations}.
\newblock \emph{\bibinfo{journal}{Physical Review E}}
  \textbf{\bibinfo{volume}{70}}, \bibinfo{pages}{017701}
  (\bibinfo{year}{2004}).

\end{thebibliography}

\onecolumngrid
%\appendix
\newcounter{figureS}
\setcounter{table}{0}
\setcounter{equation}{0}
\renewcommand{\theequation}{S\arabic{equation}}
\renewcommand{\thesection}{S\arabic{section}}
\renewcommand{\thesubsection}{\Alph{subsection}}
\renewcommand{\thesubsubsection}{\roman{subsubsection}}
\renewcommand{\thefigure}{S\arabic{figureS}}
\renewcommand{\thetable}{S\arabic{table}}

%%\begin{center}
\newpage
\tocless{\section*{Supplementary Information}}
\tableofcontents
%%\end{center}
	% !TEX root = sync_arXiv.tex      
 %---------------------------------------------------------------
 %Experimental Setup
 %---------------------------------------------------------------
 
\section{Experimental Methods}
\subsection{Detailed experimental setup}
\begin{figure}[htbp]
 \addtocounter{figureS}{1}
\begin{center}
\includegraphics[width=\textwidth]{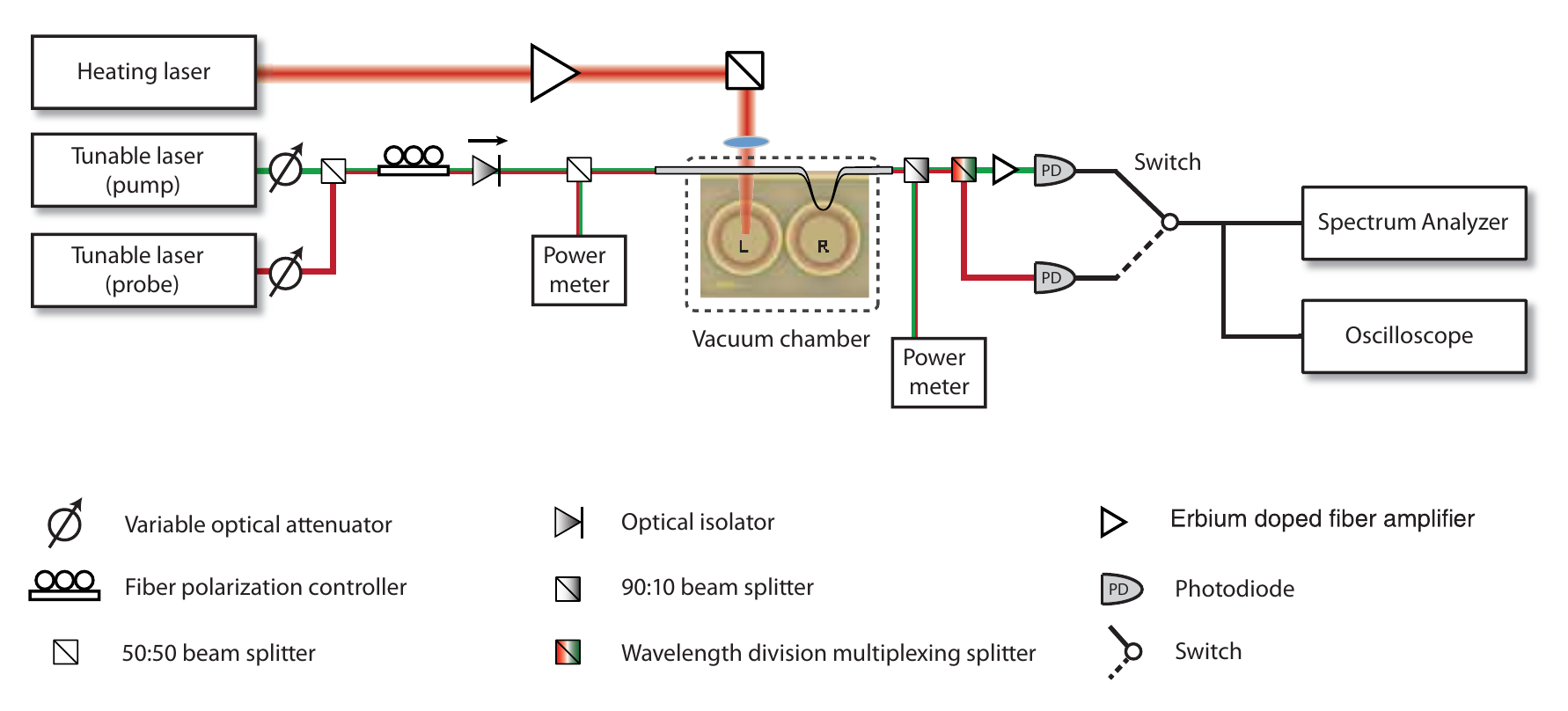}
\caption{Detailed experimental setup. See SI text for more details.}
\label{fig:setup}
\end{center}
\end{figure}

We measure the optomechanical transduction of the coupled OMOs using the setup shown in figure~\ref{fig:setup}. The green (red) line indicates the pump (probe) laser path. The probe is only used when taking the pump-probe measurements. The radio frequency (RF) spectral maps shown in the main text and in figure~\ref{fig:setup} are obtained with the probe laser off. Both the pump and the probe laser are fibre-coupled, tunable, near-infra (IR) lasers (Tunics Reference and Ando AQ4321D). Their optical power is controlled using independent variable optical attenuators. The pump and probe light are individually sent to a polarization controller and combined with a $50:50$ directional fiber coupler. A fraction of the power is monitored by a power meter which indicates the equivalent input optical power to the system. To prevent the back scattered light from entering the laser, an optical isolator is used before feeding the laser into a vacuum probe station (Lakeshore TTPX) operating at a pressure of $10^{-5}$~mT. The light is evanescent coupled to the OMOs through a tapered optical fibre waveguide by using a micro positioning system.
    
A small portion of the transmitted light ($10\%$) is also monitored by a power meter. The remaining transmitted light is split with a wavelength division multiplexing coupler to separate the pump and the probe laser. Since the pump power used is low, especially for sub-threshold measurements, the pump light is optionally amplified with a low noise erbium pre-amplifier (EDFA, Amonics AEDFA-PL-30) before coupling to a $125$~MHz bandwidth photodiode (New Focus 1181). An additional detector (Thorlabs PDB150C-AC) can be switched on when the probe measurement is necessary. Half of the detected signal is sent to an oscilloscope and the remaining is coupled to a radio-frequency spectrum analyser (RSA, Agilent E4407B).
    
The heating light source is provided by another near-IR laser (JDS SWS16101), operating at 1550 nm, and amplified by a high power EDFA (Keopsys KPS-CUS-BT-C-35-PB-111-FA-FA) that can provide a maximum power of $2$~W. The light is sent to the microscope optics which focus the light on to the device. Typically, $50$~mW of laser power is needed to achieve the desired tuning range, details of tuning aspect can be found in section ~\ref{sec:thermal_tuning}. 
  
  %-----------Single cavity measurement------------
  \subsection{Measurements}
The RF spectral maps are obtained  by detuning the laser from blue to red into the optical resonance in a stepwise fashion, as controlled by a voltage applied to laser's external cavity piezo; the laser used has a tuning coefficient of $1.1$~GHz/volt. For each voltage step, the RF spectrum is recorded. Therefore, the step size determines the vertical resolution of the RF spectra map (see Fig.3 main text) whereas the resolution bandwidth of the RSA determines the horizontal resolution. Here we used a detuning step size of $3$ MHz and a resolution bandwidth of $1$ kHz ($100$ Hz video bandwidth). This allows us to obtain a high resolution map while keeping the data collection time reasonable ($\approx20$ minutes).
   %-----------{Pump probe measurement------------ 
  \subsection{Single and coupled cavities measurement}  
The single cavity data are obtained by coupling the tapered fiber either to the $L$ or $R$ OMO. When one OMO is tested, the remaining one is heated by the heating laser with high power ($\sim 50$~mW) to ensure that they are completely decoupled.  The coupled cavity data are obtained by coupling to the $R$ OMO with the tapered fibre. In this case we use the external heating laser to fine tune the coupling so that their split spectrum is symmetric. 
  %-----------{Pump probe measurement------------ 
  \subsection{Pump probe measurement}
  The pump probe measurements provide direct evidence for the synchronization of the two OMOs. The individual probe of each cavity, as shown in Fig. 4 main text, relies on the asymmetric coupling of one the higher order optical supermodes. This asymmetry arises due to their different optical resonant frequency (See section~\ref{sec:fem_modes}) which  stems from the slight difference in the geometry of the two OMOs. This leads to a different mode splitting for the higher and lower order optical modes. In the devices we have tested, the majority of them show similar non-identical mode splitting.
  
Due to its lowers optical quality factor ($Q$) and reduced optomechanical coupling $g_{om}$, the threshold power for self-sustaining oscillations \cite{AneArcUnt0912, KipVah07} of the probe resonance is $P_{th_{probe}} \approx 20$~mW, which is roughly $20,000$~times larger than the pump resonance threshold optical power $P_{th_{pump}}\approx1~$\textmu W. We used a probe power of $P_p= (20\pm2)~$\textmu W, ensuring a low-noise detected probe signal without affecting the cavity oscillation dynamics. 
%  
%We describe the procedure for the pump-probe measurement as the following. The pump laser is turned off and the probe laser is sent to the cavity at a relative high power  $P_p\approx 200~\mu W$ so that the peaks due to amplified Brownian motion of the two OMOs can be observed on the RSA. Since the optical resonance of the probe light is asymmetric, depending on the probe frequency, the mechanical peak due to each OMO will appear separately. The detuning value when each peak appears is recorded. To probe the pump dynamics with minimum perturbation, we reduce the power of the probe laser so that the thermal Brownian peak due to the probe laser is close to the noise floor. The pump laser is parked to the desired detuning point corresponding to a particular dynamics. We monitored the probe signal by the RSA and move the probe frequency to the previously recorded frequencies to measure each OMOs contribution.
  %---------------------------------------------------------------
%OPTICAL AND MECHANICAL MODES
%---------------------------------------------------------------
\subsection{RF time-domain output power}

\begin{figure}[htbp]\begin{center}
 \addtocounter{figureS}{1}
\includegraphics[scale=0.9]{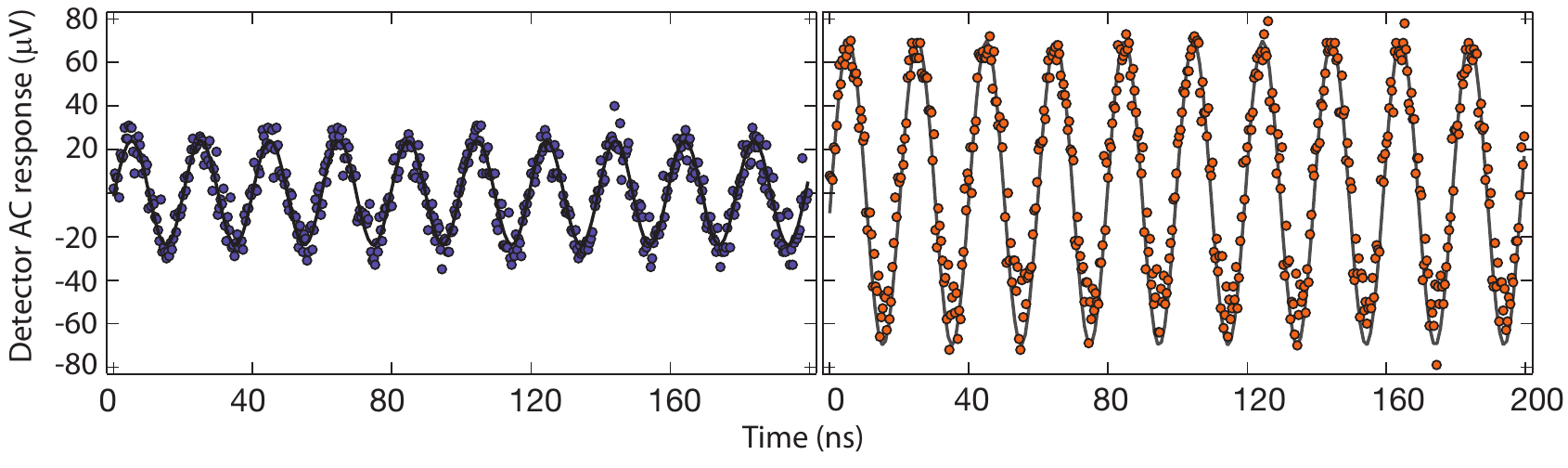}
\caption{Detector voltage time trace for the $L$ OMO oscillating only state (left) and the synchronized state (right). The synchronized optical RF power is more than 3 time higher than the $L$ OMO oscillating only state.}
\label{fig:power}
\end{center}
\end{figure}

We record the real-time trace of the output detector signal with an oscilloscope for both the asynchronous state (between dashed and solid line in Fig 3c) and the synchronized state (above solid line in Fig 3c). As shown in the RF power spectrum in Fig 3d and figure \ref{fig:power} The oscillation output RF power is increased by more than 5 dB as a the two OMO synchronizes. This is expected as both OMOs contribute to the total output optical RF power.

\section{Lumped Model Parameters}
\subsection{\label{sec:fem_modes}Optical and Mechanical modes}
To obtain the optical and mechanical modes of the optomechanical disk cavity we rely on finite element simulations using COMSOL\textregistered. From these numerical simulations we derive parameters for the lumped model that describes the optomechanical dynamics, such as the effective motional mass $m_{eff}$, and the optomechanical coupling rate $g_{om}$. The optical modes are sought by solving the Helmholtz vector wave equation with an ansatz $\vec E(r,z,\phi)=\vec E(r,z)\exp(im\phi)$. In the table~\ref{tab:fem_modes} we show the mode radial electric field profile for the lowest order optical  transverse-electric ($TE$) modes. The mechanical displacement field is sought by enforcing complete cylindrical symmetry, $\vec{u}(r,\phi,z)=\vec{u}(r,z)$, the mode profiles are also shown on table~\ref{tab:fem_modes}. From the sought eigenmodes, the optomechanical coupling coefficients for the supported optical modes are calculated using boundary perturbation theory \cite{EicChaSaf0910,JohIbaSko0206},
\begin{equation}
g_{om}\equiv\ \frac{\partial \omega}{\partial x}=\frac{\omega_{0}}{2}\frac{\int\left(\vec{U}\cdot\hat{n}\right)\left(\Delta\epsilon_{12}\left|\vec{E}\cdot\hat{t}\right|^{2}+\Delta\epsilon_{12}^{-1}\left|\vec{D}\cdot\hat{n}\right|^{2}\right)dA}{\int \epsilon \left|\vec{E}\right|^2dV},
\label{eq:gom}
\end{equation}
where the dimensionless displacement field is defined as $\vec{U}\equiv \vec{u}/\max\left|\vec{u}\right|$, the relative permitivity differences are given by $\Delta\epsilon_{12}=\epsilon_1-\epsilon_2$ and $\Delta\epsilon_{12}^{-1}=1/\epsilon_1-1/\epsilon_2$, the unit vectors $\hat{t}$ and $\hat{n}$ indicate the tangential and normal components of the vectors. The effetive motional mass is calculated as,
\begin{equation}
m_{eff}=\int\rho\left|\vec{U}\right|^{2}dV.
\end{equation}

\begin{table}[h!]
\centering
\begin{minipage}{0.5\textwidth}
\centering
\includegraphics[scale=.70]{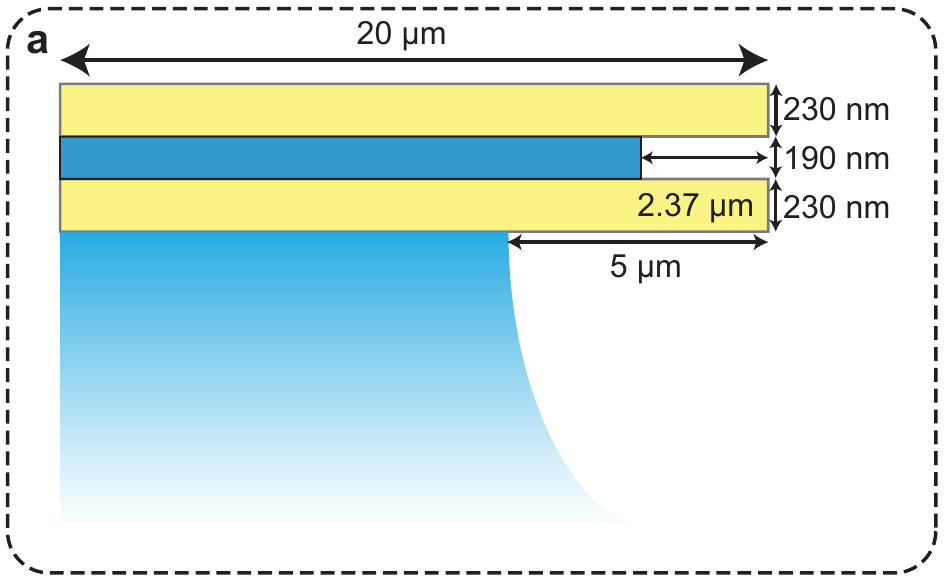}
\label{fig:fig1}
\vspace{10pt}
\begin{tabular}{|c|c|c|}
\hline 
Mechanical mode & $\frac{\Omega_{m}}{2\pi}$ (MHz) & $m_{eff}$ (pg)\tabularnewline
\hline
\includegraphics[width=3cm]{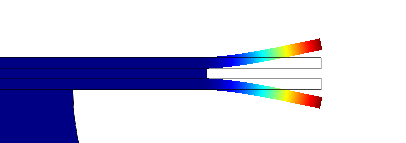} & 50.5 & 110\tabularnewline
\hline 
\includegraphics[width=3cm]{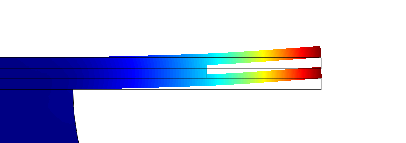} & 28.7 & 194\tabularnewline
\hline 
\end{tabular}
\end{minipage}
%~\hfill~
\\
\vspace{10pt}
\begin{minipage}{0.65\textwidth}
\centering
\begin{tabular}{|c|c|c|c|}
\hline 
Profile ($|\vec{E}\cdot\hat{r}|$) & Mode $TE_{m}^{n}$ & $\lambda_{0}$ (nm) &  $g_{om}/2\pi$(GHz/nm)\tabularnewline
\hline 
\includegraphics[width=3cm]{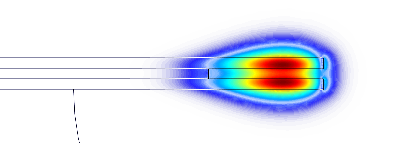} & $TE_{115}^{1}$ & 1582.28 &   49.4\tabularnewline
\hline 
\includegraphics[width=3cm]{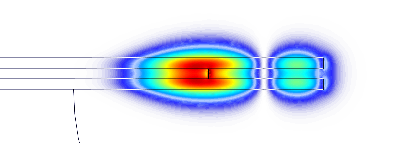} & $TE_{110}^{2}$ & 1584.87 &   11.3\tabularnewline
\hline 
\includegraphics[width=3cm]{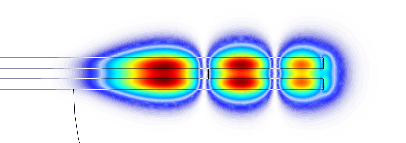} & $TE_{106}^{3}$ & 1582.31 &   17.9\tabularnewline
\hline 
\includegraphics[width=3cm]{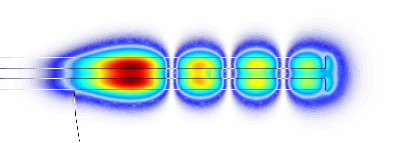} & $TE_{101}^{4}$ & 1591.01 &   10.6\tabularnewline
\hline 
\end{tabular}
\label{tab:tab1}
\end{minipage}
\caption{\label{tab:fem_modes}Optical and mechanical modes parameters. (\textbf{a}) Geometry of the optomechanical cavity used to calculate the modes and parameters shown in the tables. For the optical modes profiles, it is shown the modulus of the radial electric field $|\vec{E}\cdot\hat{r}|$; $g_{om}$ is calculated using Eq.~\eqref{eq:gom}. whereas for the mechanical modes it is shown the displacement amplitude $|\vec{u}|$ as colors and the deformation represents the normalized displacement.}
\end{table}
\subsection{\label{sec:thermal_tuning}Top illumination thermal tuning }
\begin{figure}[htbp]\begin{center}
 \addtocounter{figureS}{1}
\includegraphics[scale=0.5]{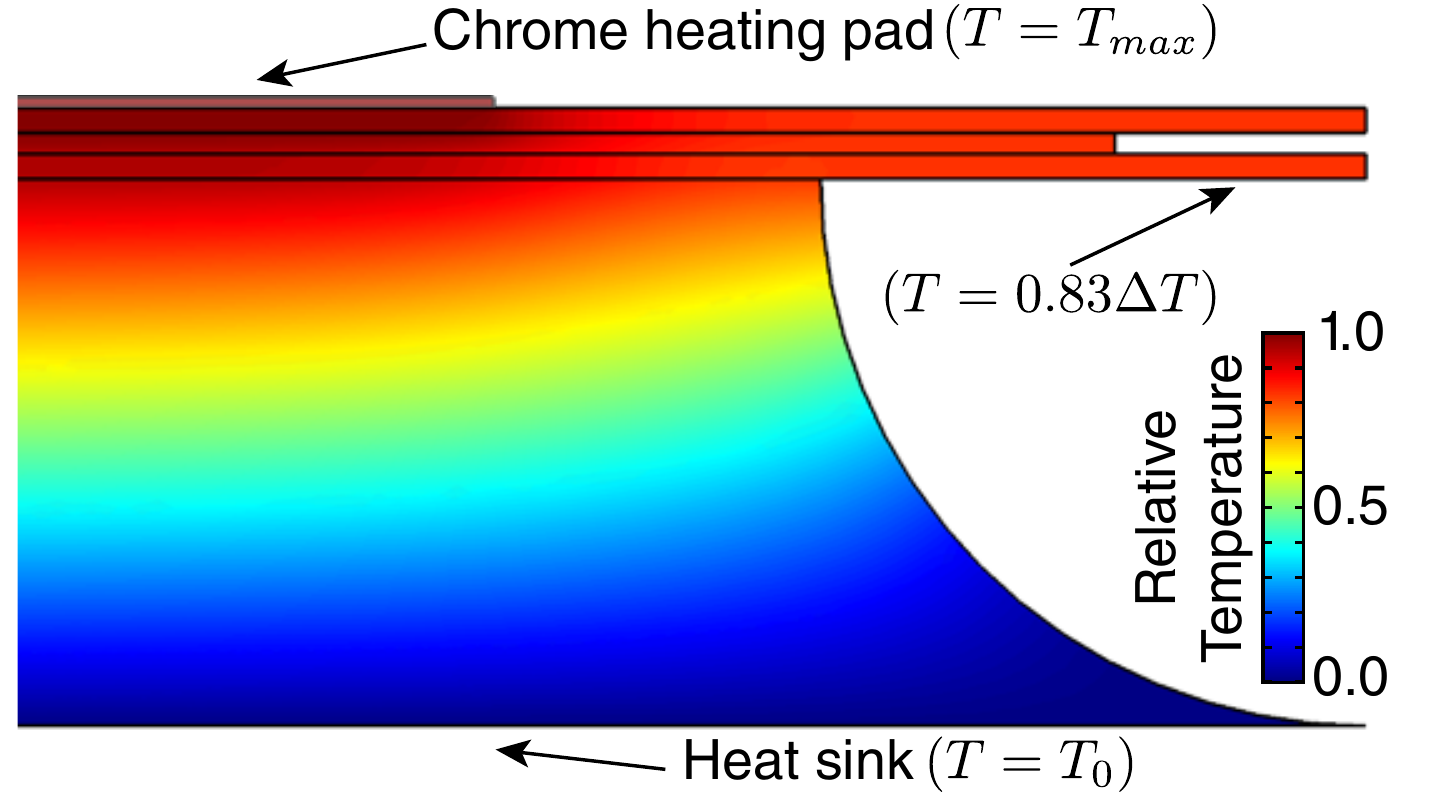}
\caption{Thermal tuning of optical resonances. Simulated temperature ($\Delta T=T-T_0$) profile of the optical micro cavity. The bottom boundary act as a heat reservoir with constant temperature $T_0=300$ K. In the mirroring edge, where the optical modes are localized, the temperature  is $T\approx0.83 \Delta T$ }
\label{fig:temperature}
\end{center}
\end{figure}

 The coupling between the cavities is controlled by changing their resonant frequencies through the thermo-optic effect. We choose to use 200 nm thick chrome pads as the heating element since they absorb $25 \%$ of 1550 nm light at normal incidence, taking into account its reflectivity. Chrome is also resistant to buffered oxide etch which follows in the fabrication steps.  The 1550 nm laser is amplified with an EDFA, coupled to the imaging microscope and focused on the chrome pads.  The heat absorbed by the chrome pads induces a temperature change $\Delta T= R_{th} P_{abs}$, where $R_{th}=\partial \Delta T/\partial P_{abs}\approx5.2\times 10^3$ K/W is the simulated effective thermal resistance of our device.  Due to thermo-optic effect, the temperature frequency shift rate is given by the perturbation expression,
\begin{equation}
g_{th}=\frac{\partial\omega_T}{\partial \Delta  T}=-\frac{\omega_0}{2 n_g}\frac{\int \alpha(r,z)T_{rel}(r,z)|\vec{E}|^2 dV}{\int |\vec{E}|^2dV}
\label{eq:g_th}
\end{equation}
where $0<T_{rel}(r,z)<1$ is the dimensionless relative temperature distribution of the device, $\alpha$ is the material-dependent thermo-optic coefficient, and  $n_g$ is the optical mode group index. If we define the overlap integral $\Gamma=\int_{SiN} |\vec{E}|^2/\int_{all} \vec{E}|^2$, Eq.~\eqref{eq:g_th} is approximately given by $g_{th}\approx-(T_{rel}^{(\text{SiN})})\omega_0\alpha_{\text{SiN}}\Gamma/(2n_g)$. In Fig.~\ref{fig:temperature} we show the simulated relative temperature field $T_{rel}(r,z)$, at the edge of the disk $T_{rel}=T_{rel}^{(\text{SiN})}\approx0.83$. From these results we can estimate the top illumination laser power needed to tune the cavity's optical frequency by $ \Delta \omega _T$,
\begin{equation}
P_{abs}= \frac{\Delta\omega_T}{g_{th}R_{th}}\approx \frac{2 n_g}{R_{th} (T_{rel}^{(\text{SiN})})\alpha_{\text{SiN}}\Gamma}\left(\frac{\Delta\omega_T}{\omega_0}\right)
\label{eq:pabs}
\end{equation}
For our device, tuning of $\delta \lambda\approx0.2$ nm is sufficient to completely decouple the two cavity modes. Using $n_g\approx1.8$, $\alpha_{\text{SiN}}=3\times10^{-5} \text{ K}^{-1}$, and $\Gamma\approx0.59$, Eq.~\eqref{eq:pabs} gives a tuning efficiency   $g_{th}/2\pi\approx-256$  MHz/K, therefore a laser power of $P= P_{abs}/25\%\approx24$ mW is needed to control the optical coupling between the cavities (see section~\ref{sec:transmission}).This value is in reasonable agreement with the experimental power range.
%
%\begin{table}
%\begin{centering}
%\begin{tabular}{|c|c|c|}
%\hline 
%Mechanical mode & $\frac{\Omega_{m}}{2\pi}$ (MHz) & $m_{eff}$ (pg)\tabularnewline
%\hline
%\includegraphics[width=3cm]{mode_mech_50mhz_crop.png} & 50.5 & 110\tabularnewline
%\hline 
%\includegraphics[width=3cm]{mode_mech_29mhz_crop.png} & 28.7 & 194\tabularnewline
%\hline 
%\end{tabular}
%\par\end{centering}
%\caption{\textbf{Mechanical modes}.}
%\end{table}

 %---------------------------------------------------------------
%COUPLED CAVITY DYNAMICS
%---------------------------------------------------------------
% \section*{Coupled Cavities dynamics}
% \begin{figure}[htbp]
%\begin{center}
%\includegraphics[scale=1.0]{figures/figure_S1.pdf}
%\caption{\textbf{Schematic of the optical and mechanical degrees of freedom}. (a) Two optical double-disk optomechanical resonators are coupled through the optical near-field with a coupling strength $\kappa/2$, $\gamma_{i_{1,2}}$ represents the intrinsic loss of each cavity and  $\gamma_{e}$ is the coupling rate of cavity 1 to the bus waveguide. (b) The mechanical degrees of freedom of each optomechanical resonator, represented by the two pendula,  are \textit{not} mechanically coupled, however, because of the optical coupling they talk to each other through the mechanically modulated photons shared among the two cavities.}
%\label{fig:coupled_rings}
%\end{center}
%\end{figure}
\section{Coupled Optomechanical Oscillators}
\subsection{Coupled mode equations}
The optical modes $a_1$ and $a_2$ of each optical cavity are coupled through the optical near-field. Due to scattering, there is also coupling between the clockwise $(cw)$ and counter-clockwise $(ccw)$ optical modes, therefore we need to consider four optical modes,  $a_{1}^{(cw,ccw)}$ and  $a_{2}^{(cw,ccw)}$. The coupled equations satisfied by these modes are given by \cite{Hau84,GorPryIlc0006} ,
\begin{equation}
\begin{aligned}
\label{eq:opt_diagonal}
\left(\begin{array}{c}
\dot{a}_{1}^{cw}\\
\dot{a}_{1}^{ccw}\\
\dot{a}_{2}^{cw}\\
\dot{a}_{2}^{ccw}
\end{array}\right)&=
\left(
\begin{array}{cccc}
 -\frac{\gamma _1}{2}-\mathrm{i} \omega _1 & \frac{\mathrm{i} \beta }{2} & \frac{\mathrm{i} \kappa }{2} & 0 \\
 \frac{\mathrm{i} \beta }{2} & -\frac{\gamma _1}{2}-\mathrm{i} \omega _1 & 0 & \frac{\mathrm{i} \kappa }{2} \\
 \frac{\mathrm{i} \kappa }{2} & 0 & -\frac{\gamma _2}{2}-\mathrm{i} \omega _2 & \frac{\mathrm{i} \beta }{2} \\
 0 & \frac{\mathrm{i} \kappa }{2} & \frac{\mathrm{i} \beta }{2} & -\frac{\gamma _2}{2}-\mathrm{i} \omega _2
\end{array}\right)
\left(\begin{array}{c}
{a}_{1}^{cw}\\
{a}_{1}^{ccw}\\
{a}_{2}^{cw}\\
{a}_{2}^{ccw}
\end{array}\right)+\sqrt{\gamma _1 \eta _{c}}s_1(t) 
\left(\begin{array}{c}
1\\
0\\
0\\
0
\end{array}\right)
\end{aligned}
\end{equation}
where $\omega_m$ are optical resonance angular frequencies, $\gamma_m$ is total damping rate, $\kappa/2$ is the  inter-cavity optical coupling rate, and $\eta_c=\gamma_e/(\gamma_{i_1}+\gamma_e)$ is the coupling ideality factor, where $\gamma_e$ is the external loss rate (due to the bus waveguide) and $\gamma_i$ is the intrinsic damping rate\cite{SpiKipPai0307}.

The system of Eqs.~\eqref{eq:opt_diagonal} can be diagonalized exactly, each eigenvector is governed by an equation of the form
\begin{equation}
\label{eq:eigvector1}
\dot{b}_{(m,\pm)}=\left[-\mathrm{i}\left(\bar{\omega} +(-1)^m\xi/2 \pm \beta/2\right)-\bar{\gamma}/2\right]b_{(m,\pm)}
(\pm)^m \frac{\kappa  \sqrt{\gamma _1 \eta _{c}}s_1(t)  }{2 \xi },\text{ for }m=1,2,
\end{equation}
where  $\bar{\omega}=(\omega_1+\omega_2)/2$, $\bar{\gamma}=(\gamma_1+\gamma_2)/2$ and   $\xi=\kappa\sqrt{1-(\delta/\kappa)^2}$, where $\delta=(\gamma_1-\gamma_2)/2+\mathrm{i}(\omega_2-\omega_1)$. The original fields $a_{1,2}^{cw,ccw}$ can be recovered from the eigenvectors through the relation,
\begin{equation}
\label{eq:eig1}
\left(\begin{array}{c}
a_{1}^{cw}\\
a_{1}^{ccw}\\
a_{2}^{cw}\\
a_{2}^{ccw}
\end{array}\right)=
\frac{1}{2\xi}\left(
\begin{array}{c}
- \kappa  b_{(1,-)}+(\xi +i \delta ) b_{(2,-)} \\
 -\kappa b_{(1,+)} +(\xi +i \delta ) b_{(2,+)} \\
 \kappa  b_{(1,-)}+(\xi -i \delta ) b_{(2,-)} \\
 \kappa  b_{(1,+)}+(\xi -i \delta ) b_{(2,+)}
\end{array}
\right)
\end{equation}
where $b_{(m,\pm)}=\left(a_m^{ccw}\pm a_m^{cw}\right)$, Eq.~\eqref{eq:eig1} will be used to calculate the optical transmission function in the section~\ref{sec:transmission} below.
%---------------------------------------------------------------
%OPTICAL TRANSMISSION
%---------------------------------------------------------------
%\section*{Optical transmission}

%-----------steady state transmission------------
\subsection{\label{sec:transmission}Steady-state transmission}
\begin{figure}[htbp]
 \addtocounter{figureS}{1}
\begin{center}
\includegraphics[scale=1.0,angle=90]{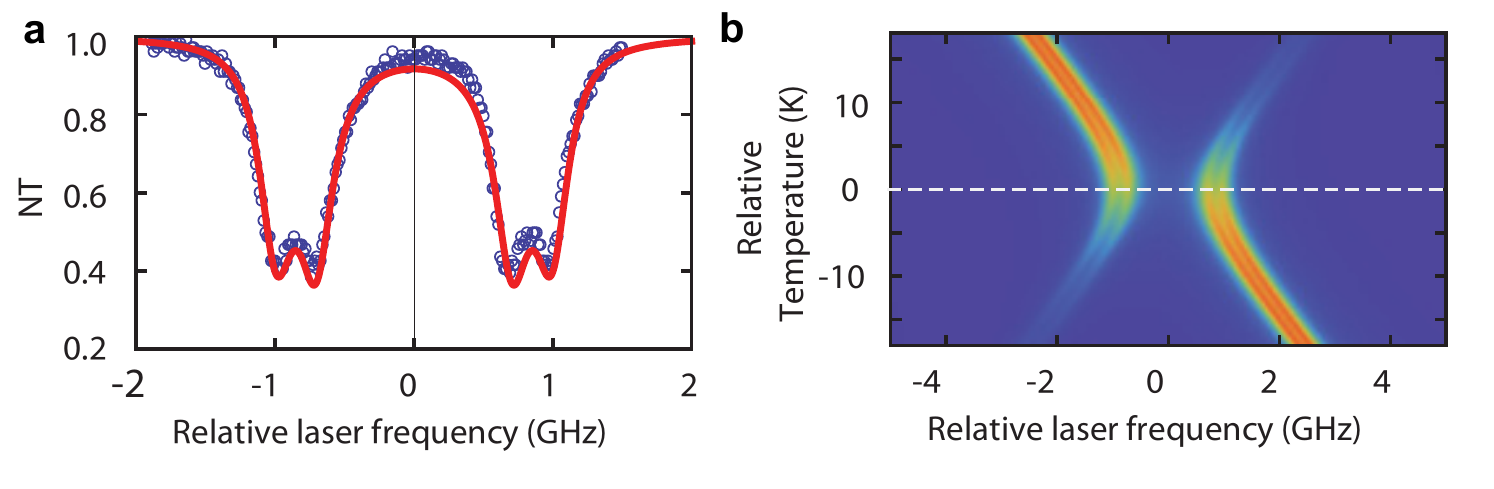}
\caption{\textbf{Optical transmission}. (\textbf{a}) Best-fit steady-state normalized optical transmission (red-line), calculated using equation~\eqref{eq:opt_trans_full}, and measured transmission spectrum (blue circles). The fit parameters are described in the text. (\textbf{b}) Optical transmission showing the thermal tuning of the coupled cavities, the false-color scale indicates the transmission. This map is obtained from~\eqref{eq:opt_trans_full} using $\omega_1(T)=\omega_{1_0}+g_{th}\Delta T$, in good agreement with Fig. 2 in the main text.}
\label{fig:splitting}
\end{center}
\end{figure}

To obtain the low-power steady-state optical transmission spectrum, we assume that the laser driving term in Eq.~\eqref{eq:opt_diagonal} is oscillating at $\omega$, i.e., $s_1(t)=s_1 e^{\mathrm{i}\omega t}$. Eq.~\eqref{eq:eigvector1} can be written in a rotating frame $c_{(m,\pm)}(t)=\tilde{c}_{(m,\pm)}(t)e^{\mathrm{i}\omega t}$.  The resulting equations will be of the form,
\begin{equation}
\label{eq:eigvector_rotating}
\dot{\tilde{b}}_{(m,\pm)}=\left[\mathrm{i}\Delta_{(m,\pm)}-\bar{\gamma}/2\right]\tilde{b}_{(m,\pm)}
(\pm)^m \frac{\kappa  s_1 \sqrt{\gamma _1 \eta _{c}}}{2 \xi },\text{ for } m=1,2,
\end{equation}
where $\Delta_{(m,\pm)}=\omega-\left(\bar{\omega} +(-1)^m\xi/2 \pm \beta/2\right)$ is the laser-cavity frequency detuning for each of the optical supermodes. The steady-state solution to~\eqref{eq:eigvector_rotating} is given by
\begin{equation}
\label{eq:eigvector_rotating_sol}
\tilde{b}_{(m,\pm)}=(\mp)^m \frac{\kappa  s_1 \sqrt{\gamma _1 \eta _{c}}}{2 \xi  \left[\mathrm{i}\Delta_{(m,\pm)}-\bar{\gamma}/2\right]},\text{ for }m=1,2.
\end{equation}
The driving laser excites directly only the mode ${a}_{1}^{cw}$, therefore the steady state optical field transmitted through the bus waveguide is given by,
\begin{equation}
s^{out}_1(\omega_l)=s_{1}-\sqrt{\gamma_1 \eta_c}{a}_{1}^{cw}
\label{eq:opt_trans}
\end{equation}
where the optical field $a_1(\omega_l)$ is given by Eq.~\eqref{eq:eig1}. The normalized field transmission,  $t(\omega)=s^{out}_1(\omega)/s_1$ is given by,
\begin{equation}
%\begin{multline}
t=1-\mathrm{i} \frac{\gamma _1\eta_c \kappa }{2\xi^2}\sum_{j=1,2} \left(\frac{\xi +\mathrm{i} \alpha  \kappa }{(-1)^{j}\beta +\xi+2 \bar{\Delta} +\mathrm{i} \bar{\gamma} }+\frac{(-1)^{j}\kappa }{(-1)^{j}\beta-\xi+2 \bar{\Delta} +\mathrm{i}\bar{ \gamma}  }\right),
\label{eq:opt_trans_full}
%\end{multline}
\end{equation}
where $\bar{\Delta}=\omega_l-\bar{\omega}$ is the detuning from the average frequency of the two cavities. The normalized power transmission is obtained from the relation $T(\omega)=\left|t(\omega)\right|^2$. In Fig.~\ref{fig:splitting} we show the transmission $T(\omega)$ using the best-fit parameters $\bar{\omega}/(2\pi)=188.442$~THz, $\bar{\gamma}/2\pi=299$~MHz,  
 $(\kappa,\beta)/2\pi=(1700,298)$~MHz, and $\eta_c=0.65$. The fit loaded optical quality factor is $Q=\bar{\omega}/\bar{\gamma}=630,000$. To obtain the thermal tuned transmission of our device, we use Eq.~\eqref{eq:opt_trans_full} together with the results described in section~\ref{sec:thermal_tuning}. The resonant frequency of the cavities, when the top-illumination is on, is given by is given by $\omega_m(T)=\omega_{m_0}+g_{th} \Delta T$, where $g_{th}/2\pi \approx -256$  MHz/K (see section~\ref{sec:thermal_tuning}).
%--------%
\subsection{\label{sec:int_mech}Intrinsic mechanical frequencies}
\begin{figure}[htbp]
 \addtocounter{figureS}{1}
\begin{center}
\includegraphics[scale=1.0,angle=0]{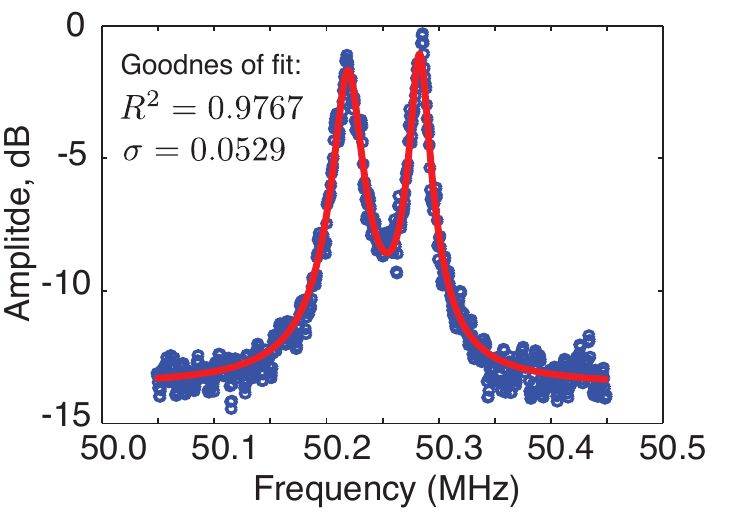}
\caption{\textbf{Mechanical modes RF spectrum}. (\textbf{a}) Double-Lorentzian best-fit steady-state normalized RF spectrum (red-line), and measured RF spectrum (blue circles). The fit parameters are described in the text.}
\label{fig:mechanical_fit}
\end{center}
\end{figure}

Using the optical read-out of the mechanical motion, as described in section \ref{sec:transduction}, we measured the mechanical quality factors using a low power optical probe coupled to the lower frequency optical supermodel (right peak in Fig. \ref{fig:splitting}a). The RF spectrum (100 averages) showing the two mechanical modes is shown in Fig. \ref{fig:mechanical_fit}. The fit parameters are the mechanical frequencies and quality factors: $(f_L,f_R)=(50.283,50.219)$~MHz and $(Q_{m_L},Q_{m_R})=(3.4\pm0.3, 2.3\pm0.2)\times 10^3$. Note that these intrinsic frequencies are slightly lower than the OMO self-sustaining oscillation frequency. This is due to the optical spring effect explained in the main text.

%-----------Mechanical equations and optomechanical coupling------------
\subsection{Mechanical equations and optomechanical coupling}

The mechanical degrees of freedom of each cavity $x_1,x_2$ follows the usual optomechanical equations \cite{Law9503,EicCamCha0905,KipVah07,CarRokYan05},
\begin{subequations}
\label{eq:mec_osc_a}
\begin{align}
\ddot{x}_1=-\Gamma_1 \dot{x}_1-\Omega_1^2 x_1+\frac{g_{om}}{m_{eff}^{(1)}\omega_0}\left(\left|a_1^{cw}\right|^2+\left|a_2^{ccw}\right|^2\right)+F^T_1(t),\\
\ddot{x}_2=-\Gamma_2\dot{x}_2-\Omega_2^2 x_2+\frac{g_{om}}{m_{eff}^{(2)}\omega_0}\left(\left|a_1^{cw}\right|^2+\left|a_2^{ccw}\right|^2\right)+F^T_2(t),
\end{align}
\end{subequations}
where $\Omega_i,\Gamma_i,m_{eff}^{(i)}$ represent the mechanical resonant frequency, dissipation rate, and effective motional mass. $F_T(t)$ is the thermal Langevin random force with expectation value $\left<F_i^T\right>=0$ and correlation function $\left<F_i^T(t)F_i^T(t+\tau)\right>=2k_B T m_{eff}^{(i)}\Gamma_i\delta(\tau)$, where $k_B$ is the Boltzmann constant and $\delta(\tau)$ is the Dirac delta function. In contrast to the phonon-laser regime\cite{GruLeePai1002}, we ignore terms which couples, through the mechanical displacement field, the optical modes $b_{(\pm,1)}$ with $b_{(\pm,2)}$; this is justified because $\kappa\gg\Omega_{L,R}$. Here we used the optical force as the positive gradient of the energy, this is a convention but must be consistent with whether the cavity frequency increases or decreases with increasing mechanical displacement; in our case the optical frequency decreases with the mechanical displacement \cite{EicChaCam0911}.

The full optomechanical dynamics is obtained by solving simultaneously Eqs.~\eqref{eq:mec_osc_a} and~\eqref{eq:opt_diagonal}, such dynamics is discussed in detail in section~\ref{sec:sync_sim}. It is however instructive to analyze how a prescribed mechanical motion of the two mechanical oscillators is read-out through the optical modes (see section \ref{sec:transduction}), also how the optical force term in Eqs.~\eqref{eq:mec_osc_a} couples to the two of them (see section ).

%-----------transduction of mechanical motion------------
\subsection{\label{sec:transduction}Optical transduction of mechanical oscillations}
To account for the mechanical effect on the optical transmission we first assume that the mechanical motion is independent of the optical fields \cite{MarHarGir0603}, which is equivalent to ignoring the dynamical back-action. Therefore we can use Eqs.~\eqref{eq:eigvector1} for the optical eigenvectors and simply replace the optical cavity's resonant frequency by $\omega_{i}\rightarrow \omega_{i}+g_{om}x_{i}$, where $x_i$ is the mechanical displacement amplitude for each cavity. The resonant frequency of each eigenmode $b_{(m,\pm)}$ will be given by,
\begin{subequations}
\label{eq:opt_freq_mech}
\begin{align}
\omega_{(1,\pm)}(x_i,x_j)=\bar{\omega}(x_i,x_j)  \pm\xi(x_i,x_j)/2 \pm \beta/2,\\
\omega_{(2,\pm)}(x_i,x_j)=\bar{\omega}(x_i,x_j)  \pm\xi(x_i,x_j)/2 \pm \beta/2,
\end{align}
\end{subequations}
where $\bar{\omega}(x_i,x_j) =\left[\omega_i(x_i)+\omega_j(x_j)\right]/2$,  $\xi(x_i,x_j)=\kappa\sqrt{(1-[\delta(x_i,x_j)/\kappa]^2}$ and $\delta(x_i,x_j)=(\gamma_i-\gamma_j)/2+[\omega_j(x_j)-\omega_i(x_i)]$. Due to the nonlinear $\xi(x_i,x_j)$ dependence on the mechanical displacement amplitudes $x_{1,2}$, The usual analytical approach to derive the optomechanical transduction coefficient does not apply\cite{MarHarGir0603}. However we can get insight into the problem if we consider the strong optical coupling limit, i.e., $\delta(x_i,x_j)/\kappa=g_{om}(x_i-x_j)/\kappa\ll1$ which means that the optical frequency splitting between the cavities is large compared to the mechanically induced frequency shift, therefore $\xi(x_i,x_j)\approx\kappa+\mathcal{O}(\delta^2/\kappa^2)$. To further simplify the analysis we assume that the two cavities share identical optical optical properties, i.e.,  $\omega_1(x_1=0)=\omega_2(x_2=0)=\omega_0$ and $\gamma_1=\gamma_2=\gamma_0$. In this case Eq.~\eqref{eq:opt_freq_mech} is approximated by,
\begin{equation}
\label{eq:opt_freq_mech2}
\omega_{(m,\pm)}(x_1,x_2)\approx \omega_{0(m,\pm)}+g_{om}\left(x_{1}+x_{2}\right)
\end{equation}
where $\omega_{0(m,\pm)}=\omega_{0}+(-1)^{m+1}\kappa/2\pm \beta/2$. Combined with the above relations, Eq.~\ref{eq:eigvector1} yields the following equation for the optical eigenmodes $b_{(m,\pm)}$,  
\begin{equation}
\label{eq:eigvector_sideband}
\dot{b}_{(m,\pm)}=\left[-\mathrm{i}\omega_{0(m,\pm)} -\mathrm{i}g_{om}(x_1+x_2)-\bar{\gamma}/2\right]b_{(m,\pm)}
(\pm)^m \frac{\sqrt{\gamma _1 \eta _{c1}}s_1 e^{\mathrm{i}\omega t}  }{2 },\text{ for i=1,2.}
\end{equation}
%\subsection{Mechanically induced sidebands}
The equations above~\eqref{eq:eigvector_sideband} can be formally integrated for a prescribed mechanical motion $(x_i=A_i \sin (\Omega_i t+\phi_i))$.  The homogeneous solutions  ($s_1=0$) decay exponentially and does not contribute after the initial transients. To find a particular solution satisfying~\eqref{eq:eigvector_sideband} we employ a common approach relying on the Jacobi-Anger expansion\cite{MarHarGir0603,HolMeaMil1105},
\begin{equation}
\begin{aligned}
&\exp\left[{\mathrm{i}\mu_1 \cos (\Omega_1 t+\phi_1)+\mathrm{i} \mu_2\cos( \Omega_2 t+\phi_2)}\right]=\sum_{p,q=-\infty}^{\infty}{i^{p+q} J_p(\mu_1)J_q(\mu_2) e^{\mathrm{i} (p\Omega_1+ q\Omega_2) t+\mathrm{i}(\phi_1+\phi_2)}},
\label{eq:Jacobi-anger}
\end{aligned}
\end{equation}
where $\mu_i=g_{om}A_i/\Omega_i$ is the optomechanical modulation depth. Inserting Eq.~\eqref{eq:Jacobi-anger} in~\eqref{eq:eigvector_sideband} and solving the resulting equations gives,
\begin{equation}
\label{eq:opt_solp2}
\begin{aligned}
b_{(m,\pm)}&(t)=\frac{(\pm)^ms_1\sqrt{\gamma _1 \eta _{c1}}}{2}  e^{\mathrm{i}\left[\omega _l t+\sum_{j=1,2}\mu _j \cos \left( \Omega _j t+\phi _j\right)\right]} \sum _{p,q} \frac{ i^{p+q}  J_p\left(\mu _1\right) J_q\left(\mu _2\right) e^{\mathrm{i}  \left(p \Omega_1+q \Omega_2\right)t}}{\bar{\gamma}/2+i \left(-\Delta_{0(m,\pm)} +p \Omega _1+q \Omega _2\right)},
\end{aligned}
\end{equation}
where the sum over $m,n$ extends over $[-\infty,\infty]$, and $\Delta_{0(m,\pm)}=\omega _l-\omega_{0(m,\pm)} $. From Eq.~\eqref{eq:opt_solp2} we can clearly see the that cavity field exhibit tones at combinations of the mechanical frequencies $(m\Omega_1+n\Omega_2)$ of the two cavities.
%\subsection{Transmission}
\section{\label{sec:sync_toy} Toy model for synchronization}
In this section we derive an approximate model to describe the essential features of our coupled oscillators. Although we develop a first order linear approximation of the two coupled optomechanical oscillators, \textbf{they constitute an intrinsically a nonlinear system}, as described in detail elsewhere \cite{HolMeaMil1105}.
\subsection{\label{sec:ommc}Optically mediated mechanical coupling}
The optical force driving terms in Eqs.~\eqref{eq:mec_osc_a} can be written in terms of the diagonal modes $b_{(m,\pm)}$ from Eq.~\eqref{eq:opt_solp2} by using Eqs.~\eqref{eq:eig1}. As in section~\ref{sec:transduction}, for large optical coupling the terms are only resonant with the driving laser one at a time, therefore we can focus on the effect of a particular choice of $(m,\pm)$. To simplify the notation we use $\Delta_{0(m,\pm)} \equiv \Delta_m$ and $b_{(m,\pm)}\equiv b_m$ below. We also assume that effective motional mass of the individual oscillators are identical, i.e. $m_\text{eff}^{(j)}=m_\text{eff}$. The driving force in each oscillator is proportional to $|b_m|^2$,
\begin{equation}
\label{eq:opt_solp3}
F_\text{opt}^{(j)}=\frac{g_\text{om}}{\omega_{0(m)}}|b_m|^2=-\frac{g_\text{om}P_{in}\gamma _1 \eta _{c1}}{4\omega_{0(m)}}  \left|\sum _{p,q} \frac{ i^{p+q}  J_p\left(\mu _1\right) J_q\left(\mu _2\right) e^{\mathrm{i}  \left(p \Omega_1+q \Omega_2\right)t}}{\bar{\gamma}/2+i \left(-\Delta_m+p \Omega _1+q \Omega _2\right)}\right|^2,
\end{equation}
which contains both DC terms and oscillatory terms.  

Although our oscillators may exhibit large oscillation amplitude ($g_{om}x_{j}>\bar{\gamma}$), it is instructive to analyze the small amplitude dynamics arising for the the driving term in Eq. \eqref{eq:opt_solp3}. This treatment is entirely analogous to the one used to derive the optomechanical damping and spring effect in uncoupled OMO's \cite{KipVah07,EicCamCha0905,EicChaCam0911}. For the small amplitude oscillation, the modulation parameters are small, i.e., $\mu_i=g_{om}x_i/\Omega_i\ll 1$, therefore the Bessel functions in \eqref{eq:opt_solp3} can be approximated by their small argument limit, $J_n(\mu)\approx\frac{1}{n!}(\frac{\mu}{2})^{n}$. We neglect any terms which are quadratic in the $\mu_{1,2}$, which also account for summing  Eq. \eqref{eq:opt_solp3} only over $p,q=0,\pm1$ since higher order terms will result in terms which are $\mathcal{O}(\mu^2)$. The $p,q=0$ terms result in a DC component of the force,
\begin{equation}
\label{eq:opt_force_DC}
F_{\text{opt}_{DC}}^{(j)}=\sum_{j=L,R}{\frac{g_\text{om}}{\omega_{0(m)}}\left(\frac{P_{in}\gamma _1 \eta _{c1}}{\Delta_m^2+(\bar{\gamma}/2)^2}\right)}.
\end{equation}
The impact of the DC term above is to shift the static equilibirum position of the mechanical oscillators. As a result, the actual optical detuning is also shifted, to account for this DC shift we substitute $\Delta_m\rightarrow\Delta_m'$, where $\Delta_m'=\Delta_m+g_\text{om} (x_1+x_2)$.

When $p,q=\pm1$ the resulting terms are quadratic in $\mu_{1,2}$ and \textbf{will be neglected in this first order approximation}, therefore the lowest order AC terms are given by  combinations $(p,q)=(0,\pm1)$ and $(p,q)=(\pm1,0)$. 
\begin{equation}
\label{eq:opt_force_AC1}
F_{\text{opt}_{AC}}^{(j)}=\frac{g_\text{om}^2 P_{in}\gamma _1 \eta _{c1}}{\omega_{0(m)}} \sum_{j=L,R}{A_j\left[-\cos(\Omega_j t) f_\text{I}(\Delta_m',\Omega_j)  + \sin(\Omega_j t)f_\text{Q}(\Delta_m',\Omega_j) \right]}
\end{equation}
where the functions $f_{I,Q}(\Delta)$, which correspond to the in-phase ($\propto \sin{(\Omega_j t)}$) and quadrature of phase component ($\propto \cos{(\Omega_j t)}$) of the AC force, are given by
\begin{subequations}
\label{eq:opt_force_AC_f}
\begin{align}
f_\text{Q}(\Delta_m',\Omega_j)=\frac{4 (\frac{\bar{\gamma}}{2})  \Delta_m'}{2 \Omega _j^2 \left((\frac{\bar{\gamma}}{2}) ^4-\Delta_m'^4\right)+\Omega _j^4 \left((\frac{\bar{\gamma}}{2}) ^2+\Delta_m'^2\right)+\left((\frac{\bar{\gamma}}{2}) ^2+\Delta_m'^2\right)^3},\\
f_\text{I}(\Delta_m',\Omega_j)=\frac{2 \Delta_m' \left((\frac{\bar{\gamma}}{2}) ^2+\Delta_m'^2-\Omega_ j^2\right)}{2 \Omega_ j^2 \left((\frac{\bar{\gamma}}{2}) ^4-\Delta_m'^4\right)+\Omega_ j^4 \left((\frac{\bar{\gamma}}{2}) ^2+\Delta_m'^2\right)+\left((\frac{\bar{\gamma}}{2}) ^2+\Delta_m'^2\right)^3}.
\end{align}
\end{subequations}
We can now use the transformations ${\sin(\Omega_j t)\rightarrow x_j/A_j, \cos(\Omega_j t)\rightarrow \dot{x}_j/(A_j\Omega_j)}$ and rewrite \eqref{eq:opt_force_AC1} as
\begin{equation}
\label{eq:opt_force_AC2}
F_{\text{opt}_{AC}}^{(j)}=\frac{g_\text{om}^2 P_{in}\gamma _1 \eta _{c1}}{\omega_{0(m)}} \sum_{j=L,R}{\left[-x_j f_\text{I}(\Delta_m',\Omega_j)  + \frac{\dot{x}_j}{\Omega_j}f_\text{Q}(\Delta_m',\Omega_j) \right]}
\end{equation}
Equation \eqref{eq:opt_force_AC2} above shows that for each oscillator the driving force will have a component proportional to its displacement ($x_j$) and its velocity ($\dot{x}_j$). But there are also terms proportional to the displacement and velocity of the opposing OMO; these are the terms that couple the two OMOs and form the basis for synchronization in our system. \textbf{Note that if higher order terms were kept in the expansion of Eq. \eqref{eq:opt_solp3}, nonlinear terms would appear in Eq. \eqref{eq:opt_force_AC2}}.

Above we derived the small amplitude form of the optical forces driving our coupled oscillators, we did not use the fact that our cavities are in the so-called unresolved sideband regime where the mechanical frequencies are much smaller than the optical linewidth, i.e., $\Omega_j/\bar{\gamma}\approx0.2\ll1$. In this limit, Eqs. \eqref{eq:opt_force_AC_f} can be written as,
\begin{subequations}
\label{eq:opt_force_AC_f}
\begin{align}
f_\text{Q}(\Delta_m')\approx\frac{4 (\frac{\bar{\gamma}}{2})  \Delta_m'}{\left((\frac{\bar{\gamma}}{2}) ^2+(\Delta'_m)^2\right)^3},\\
f_\text{I}(\Delta_m')\approx\frac{2 \Delta_m' }{\left((\frac{\bar{\gamma}}{2}) ^2+(\Delta_m')^2\right)^2}.
\end{align}
\end{subequations}

Now we can write the \eqref{eq:mec_osc_a} as two coupled harmonic oscillators,
\begin{subequations}
\label{eq:mec_osc_b}
\begin{align}
\ddot{x}_1+\Gamma_1'\dot{x}_1+(\Omega_1' )^2x_1=-k_\text{I} x_2 + k_\text{Q}\dot{x}_2,\\
\ddot{x}_2+\Gamma_2'\dot{x}_2+(\Omega_2' )^2x_2=-k_\text{I} x_1 + k_\text{Q}\dot{x}_1 ,
\end{align}
\end{subequations}
where the  modified frequency and damping rate are given by (assuming $\delta\Omega_j^2\approx2\Omega_j\delta\Omega_j$),
\begin{subequations}
\label{eq:mec_osc_sd}
\begin{align}
\Gamma_j'=\Gamma_j-\beta_j f_Q(\Delta_m'),\\
\Omega_j' =\Omega_j +\frac{\beta_j}{2}f_\text{I}(\Delta_m') ,\\
k_Q^{(j)} =\beta_j \Omega_j f_Q(\Delta_m'),\\
k_\text{I}^{(j)} =\beta_j \Omega_j f_\text{I}(\Delta_m')
\end{align}
\end{subequations}
with $\beta_j=g_\text{om}^2 P_{in}\gamma _1 \eta _{c1}/(m_\text{eff}^{j}\omega_{0(m)}\Omega_j)$. 

Therefore, in the small modulation regime ($\mu_{1,2}\ll1$), our system resemble  harmonic oscillators in which both the damping and frequency are controlled by the optical field; this result is exactly what one  would get from uncoupled OMOs. With the reduction of the mechanical damping rate for a blue detuned laser. ($\Delta_m'>0$), these two damped oscillators may undergo a bifurcation when the effective damping rate ($\Gamma_i'$) reverses sign. In this first order approximation there is no additional nonlinearity to prevent the oscillations to grow unbound, however it is known that the higher order terms in the force expansion (\eqref{eq:opt_solp3}) will balance the amplitude growth and eventually lead to a stable periodic orbit (limit cycle) \cite{HeiLudQia1107,MarHarGir0603,HolMeaMil1105}. The optical coupling in our system couples the two harmonic oscillators with both amplitude and velocity dependent terms, with coupling strengths  $k_\text{I},k_Q$, respectively. The functional dependence of such coupling is the same as the self-induced optical spring and damping rate, as given by \eqref{eq:mec_osc_sd}. 

\subsection{\label{sec:kuramoto}Approximate Kuramoto model}
\begin{figure}[htbp]\begin{center}
 \addtocounter{figureS}{1}
\includegraphics[width=\textwidth]{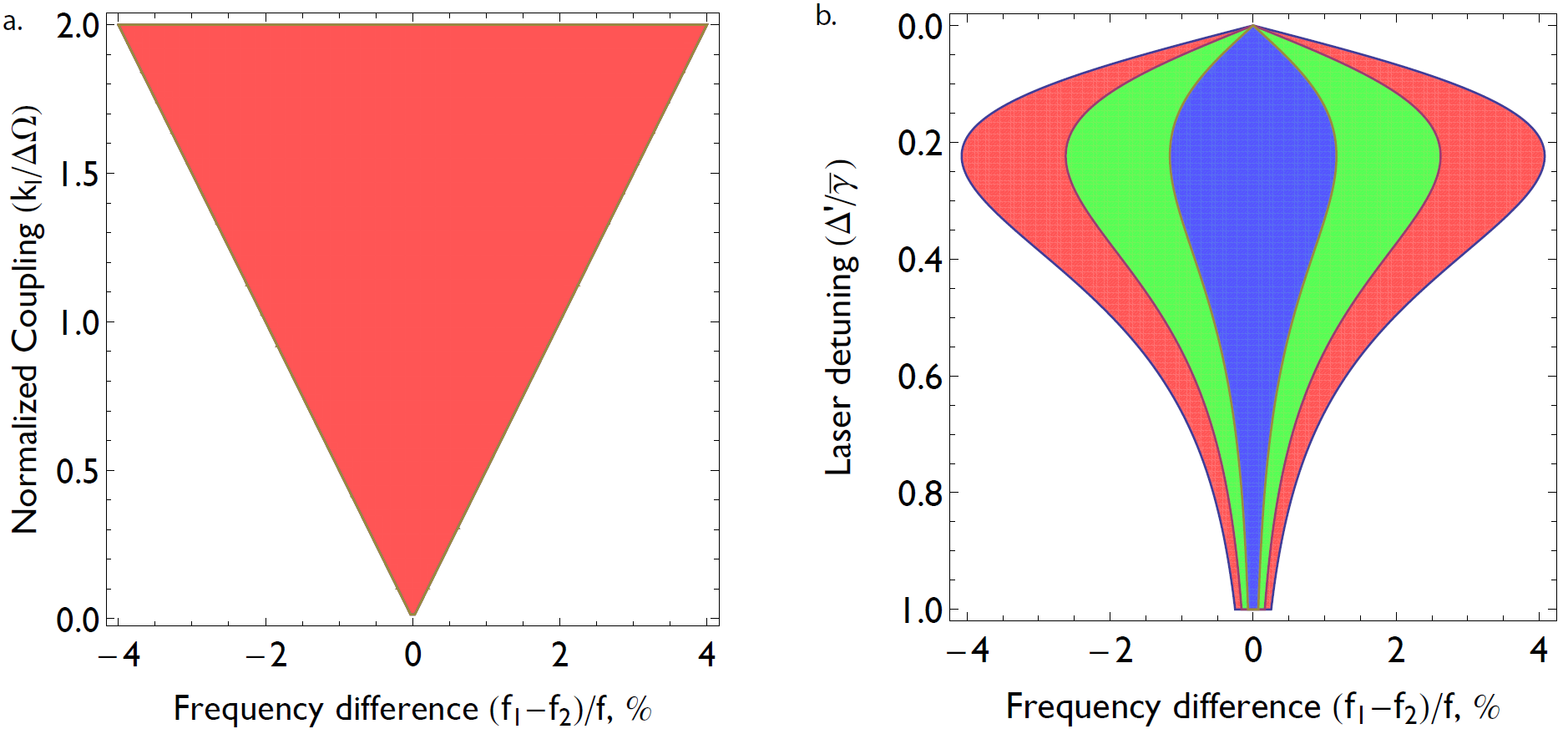}
\caption{Arnold tongue for the simplified Kuramoto model, inside the tongues the system can exhibit synchronized oscillation. (a) Usual tongue as a function of the coupling parameter $k_\text{I}$. (b) Tongue when $k_\text{I}$ is explicitly written in terms of the laser-cavity detuning. The three tongues in (b) are obtained with the optical input power values of $P_{in}={1,5, 15}$ \textmu W, the lower power is the blue whereas the highest power is red tongue. $f$ is the average mechanical frequency of the oscillators.}
\label{fig:tongue}
\end{center}
\end{figure}

In the small amplitude approximation that lead to Eq. \eqref{eq:mec_osc_b}, one can also derive slowly-varying phase and amplitude equations that describe the dynamics of our system. To accomplish this we assume the following form for our displacement amplitudes,
 \begin{subequations}
\label{eq:ku_ansatz}
\begin{align}
x_1(t)=r_1(\tau)\exp\mathrm{i}\phi_1(\tau)\exp\mathrm{i}\Omega t,\\
x_2(t)=r_2(\tau)\exp\mathrm{i}\phi_2(\tau)\exp\mathrm{i}\Omega t.
\end{align}
\end{subequations}
where $\Omega=(\Omega_1+\Omega_2)/2$ is the average frequency of oscillation. Substituting this ansatz in Eqs. \eqref{eq:mec_osc_b}  and assuming that the negative damping induced by the optical wave exactly balances the intrinsic viscosity of the oscillators ($\Gamma'\approx0$) we obtain the following amplitude and phase equations,
\begin{subequations}
\label{eq:ku_aphi2}
\begin{align}
\dot{r}_1=  -\frac{k_Q r_2}{4 \Omega ^3} \left(\left(\Omega ^2+\Omega_2^2\right) \cos (\Delta \phi )\right)-\frac{k_\text{I} r_2 \sin (\Delta \phi )}{2 \Omega },\\
\dot{r}_2=-\frac{k_Q r_1}{4 \Omega ^3} \left( \left(\Omega ^2+\Omega_1^2\right) \cos (\Delta \phi )\right)+\frac{k_\text{I} r_1 \sin (\Delta \phi )}{2 \Omega },\\
\dot{\Delta\phi}=2 \Delta \Omega+\frac{\left(R^2-1\right) \cos (\Delta \phi ) \left(\gamma  k_\text{Q}+2 k_\text{I}\right)}{2 R \Omega }+\frac{\left(R^2+1\right) \sin (\Delta \phi ) k_Q}{R}
\end{align}
\end{subequations}
where $\Delta\phi=\phi_1-\phi_2$, $R=r_1/r_2$ and $\Delta\Omega=\Omega_1'-\Omega_2'$.  Note that in this approximation the limit cycle has zero amplitude; this is the case because we neglected the higher  order terms when deriving Eq. \eqref{eq:mec_osc_b}. Despite such limitations of this toy model, the phase dynamics given by Eq. \eqref{eq:ku_aphi2} enables us to visualize how does the Arnold tongue ($|\Delta\Omega|<k_\text{I}$) behaves as we vary the laser detuning. We assume that limit cycles of the individual oscillator will have similar amplitude ($R=1$) and require  $\dot{\Delta\phi}=0$ for a synchronized oscillation. When $|\Delta\Omega|<k_\text{I}$ this conditions can be satisfied and defines an Arnold tongue for this simple model; inside the Arnold tongue the system can synchronize ($\dot{\Delta\phi}=0$).  In Fig. \ref{fig:tongue} we show the Arnold tongue plot for our system, in Fig. \ref{fig:tongue}a we plot it in the usual way, as function of the coupling coefficient ($k_\text{I}$), whereas in Fig. \ref{fig:tongue}b we plot the tongue as a function of the laser detuning by using Eq. \eqref{eq:mec_osc_sd}. Due to the model simplicity and lack of higher order terms, it does not predict the precise values for the synchronization region observed in the experiment ( Fig. \ref{fig:tongue}b), however it does agree qualitatively; for higher optical power levels the tongues get wider and allows a given mechanical frequency difference to synchronize at larger detuning, as shown in figure 3 of the main text. 

%
%
%
%If order to obtain a Kuramoto-like equation for the system obove we assume that the amplitude $r_1,r_2$ remains in steady state

%  It is interesting to note that in this approximation ($\delta/\kappa\ll1$), the output field will have only a DC term when the mechanical field amplitudes have the same modulus, frequency, but opposite phase, i.e., $A_1=-A_2$. This suggests that only in-phase synchronization should be feasible in this regime.  In our device however, as can be noted in section \ref{sec:sync_sim} below, the numerical simulations always predict out of phase synchronization. 
%\begin{subequations}
%\label{eq:ku_aphi}
%\begin{align}
%\dot{r}_1=-\frac{1}{2} \Gamma_1'  r_1-\frac{k_Q r_2}{4 \Omega ^3} \left(\Gamma'  \Omega   \sin (\Delta \phi )+ \left(\Omega ^2+\Omega_2^2\right) \cos (\Delta \phi )\right)-\frac{k_I r_2 \sin (\Delta \phi )}{2 \Omega },\\
%\dot{r}_2=-\frac{1}{2} \Gamma_2'  r_2-\frac{k_Q r_1}{4 \Omega ^3} \left(-\Gamma'  \Omega  \sin (\Delta \phi )+ \left(\Omega ^2+\Omega_1^2\right) \cos (\Delta \phi )\right)+\frac{k_I r_1 \sin (\Delta \phi )}{2 \Omega },\\
%\dot{\Delta\phi}=2 \Delta\Omega+\gamma  k_Q \left(\frac{\left(R^2+1\right) \sin (\Delta \phi )}{\gamma  R \Omega }+\frac{\left(R^2-1\right) \cos (\Delta \phi )}{2 R \Omega ^2}\right)+\frac{k_I \left(R^2-1\right) \cos (\Delta \phi )}{R \Omega }
%\end{align}
%\end{subequations}

%---------------------------------------------------------------
%Synchronization simulation
%--------------------------------------------------------------
\subsection{\label{sec:threshold}Oscillation treshold}
The individual threshold to achieve self-sustaining optomechanical oscillations can be estimated by forcing $\Gamma'=0$ in Eq. \eqref{eq:mec_osc_sd}), the resulting threshold condition ($\Delta'=\bar{\gamma}/2$) follows the expression,
\begin{equation}
P_{th}=\frac{\Omega_m m_\text{eff} \omega^4}{8 \eta_c g_\text{om}^2 Q_m Q^3}
\label{eq:threshold}
\end{equation}
where $Q$ is the optical quality factor, $Q_m$ is the mechanical quality factor, $\eta_c=\gamma_e/\gamma$ is the ideality coupling factor,  it can be obtained from the experimental spectrum through its relation to the minimum transmission value~\cite{SpiKipPai0307}. In our experiment $\eta_c=0.65$ (see Fig.~\ref{fig:splitting}). Since the threshold depends nonlinearly on several measured and calculated parameters we propagate the error through the usual relation,
\begin{equation}
\frac{\delta P_{th}}{P_{th}}=\sqrt{\left(\frac{\delta m_\text{eff}}{m_\text{eff}}\right)^2+\left(\frac{\delta \Omega_m}{\Omega_m}\right)^2+\left(4\frac{\delta \omega}{\omega}\right)^2+\left(2\frac{\delta g_\text{om}}{g_\text{om}}\right)^2+\left(3\frac{\delta Q}{Q}\right)^2+\left(\frac{\delta Q_m}{Q_m}\right)^2}
\label{eq:threshold_error}
\end{equation}
where the $\delta$ before each quantity indicates its standard deviation.  The sum of relative errors is dominated by the errors bars of $g_\text{om}$ and $Q$, using the error bars indicated in the main text we obtain $\delta P_{th}/P_{th}\approx 35\%$. This is a large deviation but is found to impact only the precise optical input power that results in a RF spectral map that matches the experimental data, as shown in the main text.  
 \section{\label{sec:sync_sim}Synchronization Simulation}
 \subsection{Simulation approach}
To simulate the synchronization dynamics and obtain the results shown in Fig. ~\ref{fig:sync_sim}, we numerically integrate the system of equations~\eqref{eq:opt_diagonal}, including the displacement dependent optical resonant frequencies, i.e. $\omega_{1,2}(x)=\omega_{1,2}+g_{om}x_{1,2}$, together with the two harmonic oscillator equations~\eqref{eq:mec_osc_a}. This is accomplished using the \textit{NDSolve} function in the commercial software Mathematica\textregistered. In the absence of the random thermal noise force in Eq.~\eqref{eq:mec_osc_a}, it is numerically challenging to capture the dynamics before the regenerative oscillation threshold is reached, this is because the steady-state is a static one, i.e., $\dot{x}_{1,2}=0$.  To overcome this issue we add a weak (low-temperature $T=1$ K) noise that prevents the dynamics to reach such static equilibrium. Since \textit{NDSolve} is a deterministic solver we include the thermal drive by assigning to $F^T_{1,2}(t)$ the outcome of a random variable with with expectation value and correlation function given by
 \begin{eqnarray}
 \label{eq:correlation}
\left<F_i^T\right> &=& 0\\
 \left<F_i^T(t)F_i^T(t+\tau)\right> &=& 2 k_B T m_{eff}^{(i)}\Gamma_i\delta(\tau),
 \end{eqnarray}
where $k_B$ is the Boltzmann constant. The discontinuity of this random driving term can lead to instabilities in $NDSolve$, to overcome this we smooth out thenoise term by interpolating the random force with a correlation time $t_c=(2\pi/\Omega_i)/30$. Such short correlation time ensures that the noise power spectrum density (PSD) is white within the frequency range of interest.The reliability of this approach is confirmed by verifying that for weak pump powers ($P\ll P_{th}$), the integrated power spectrum density $S_{x_i}(\Omega)=\left|x_i(\Omega)\right|^2$ satisfy the fluctuation-dissipation theorem \cite{SAU9010}.
\begin{equation}
\left<x^2(\Omega)\right>=\frac{1}{2\pi}\int_{0}^{\infty} S_{xx}(\Omega)d\Omega=\frac{k_B T}{2 m_{eff}^{(i)}\Omega_i^2}
\end{equation}

A complete analysis of the noise in synchronized systems is beyond the scope of this work, since an accurate numerical noise dynamics will require the simulation of the coupled non-linear stochastic dynamics of the optomechanical cavities\cite{HauJanBal0611,HoeSuaSan0109}. The computational complexity of such systems is also high due to the requirement for slow convergence, first order, fixed time step simulation \cite{BurBurHig0612,KloPla99,Wil0407}.

%Unfortunately, simulating the complete thermal noise spectrum for our device at room temperature is very computationally demanding. Since the purpose of adding noise in our simulations is to provide a constant perturbation, we reduced the mean amplitude of the thermal noise by ~4 orders of magnitude, effectively simulating the device at a very low temperature\cite{BurBurHig0612,HauJanBal0611,Hig0109,KloPla99} .
%
%We summarise the effect of this modification as the following. Reducing the noise correlation function reduces the mean of the power spectral density of the noise spectrum. This will decrease the amplitude of the spectrum before the coherent oscillation takes place. However, when the system starts to oscillate, the linewidth is no longer determined by the natural damping of the system but by the noise. The observed linewidth changing from the simulation is due to the very small linewidth of the oscillator we simulate. However it should not alter the long term dynamics of the system in the oscillating regions where the optical forces dominate.
 
 \subsection{Simulation results}
 \begin{figure}[htbp!]
  \addtocounter{figureS}{1}
 \begin{center}
 \includegraphics[scale=1.0,height=0.9\textwidth,angle=90]{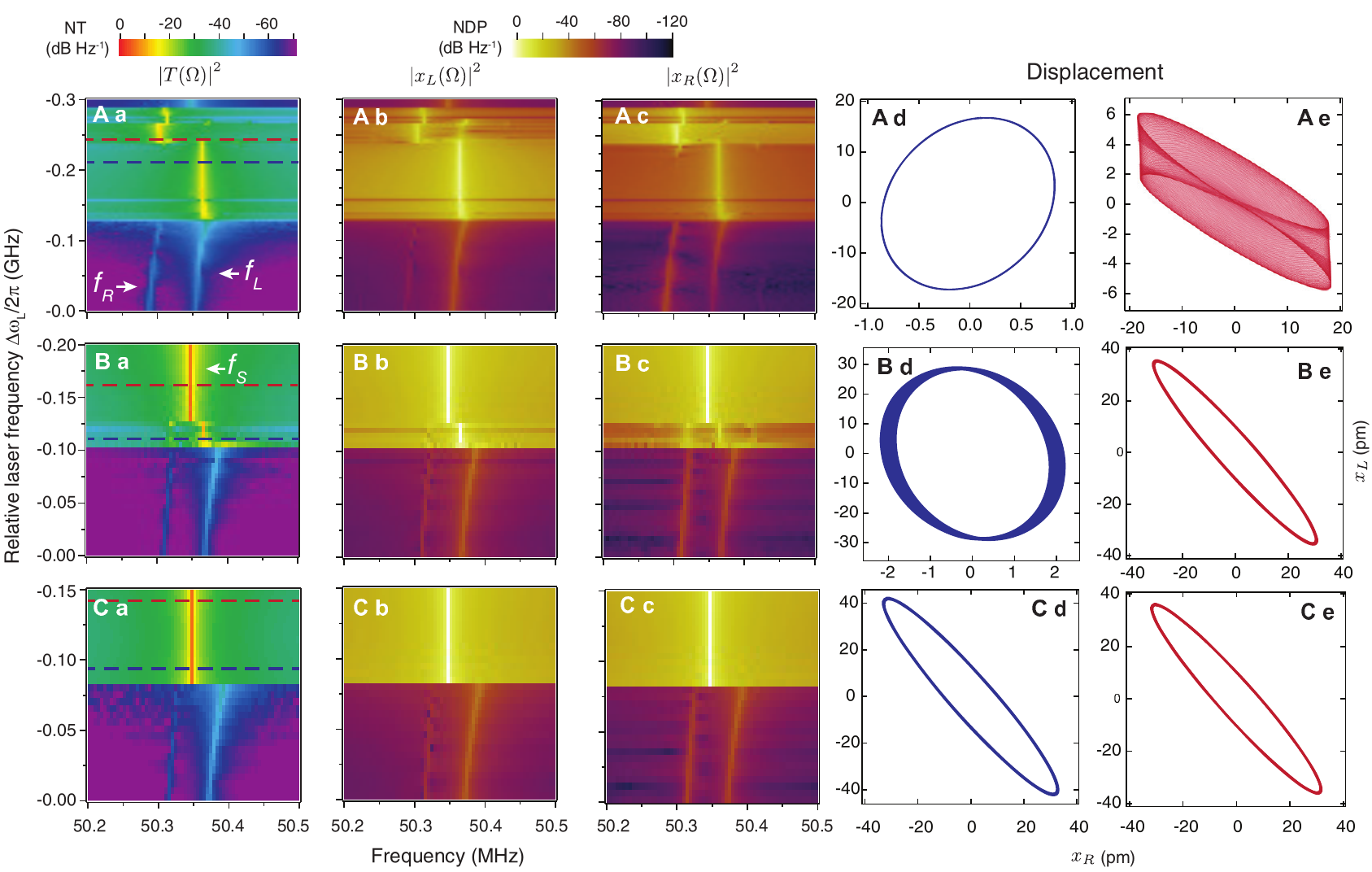}
 \caption{\textbf{Numerical simulation of the coupled oscillation dynamics}. From \textbf{a} to \textbf{e}: transmission RF spectra, displacement power RF spectra of the $L$ and the $R$ OMOs, and the displacement phase diagram of the $L$ and the $R$ OMOs, for input powers at (\textbf{A})$P_{in}=4.9$~\textmu W, (\textbf{B})$P_{in}=15.8$~\textmu W and (\textbf{C}) $P_{in}=17.9$~\textmu W. $x_L$~($x_R$): displacement of the $L$ and $R$ OMOs.}
 \label{fig:sync_sim}
 \end{center}
 \end{figure}
The simulation also allows us probe not only the optical transmission PSD, but also the mechanical displacement PSD and time series of each OMO. The complete simulation results for the pump laser powers described in the main text are shown in fig. \ref{fig:sync_sim}. The only parameter we adjusted to obtain the maps shown in figures 3 (f,g,h) in the main text and \ref{fig:sync_sim} was the optical pump power.

In figures \hyperref[fig:sync_sim]{\ref{fig:sync_sim}A} ($P_{in}=4.9$~\textmu W), the mechanical power spectrum of the oscillators (fig. \hyperref[fig:sync_sim]{\ref{fig:sync_sim}A(b,c)}) shows that for ($-0.25<\Delta\omega/2\pi<-0.13$ GHz), only $L$ OMO is oscillating; the $R$ OMO is forced to oscillate at the $L$ OMO's frequency but have not yet reached its oscillation threshold. This is illustrated by the displacement state space figures shown in fig. \hyperref[fig:sync_sim]{\ref{fig:sync_sim}A(e)} for $\Delta\omega/2\pi=-0.21$ GHz (blue dashed line in fig. \hyperref[fig:sync_sim]{\ref{fig:sync_sim}A(a)}), note that  $|x_L|$ is about 20 times larger than $|x_R|$. At $\Delta\omega/2\pi=-0.25$~ GHz, marked by the red-dashed line in fig. \hyperref[fig:sync_sim]{\ref{fig:sync_sim}A(a)}, the situation changes and the $R$ OMO oscillates with larger amplitude ($|x_R|\approx 3.5|x_L|$) but at different frequencies; the result is a Lissajous figure that fills in the whole state space.

In figures \hyperref[fig:sync_sim]{\ref{fig:sync_sim}B} ($P_{in}=15.8$~\textmu W), in the asynchronous region, indicated by the blue dashed line, the $L$ OMO oscillates with an amplitude roughly 15 times of the $R$ OMO in agreement with the measured RF spectrum and the pump probe measurement. In the unified frequency region, for both power levels $P_{in}=15.8$~\textmu W and $P_{in}=17.9$~\textmu W in fig. \hyperref[fig:sync_sim]{\ref{fig:sync_sim}C}, the phase diagram shows the two oscillators are synchronized and their amplitude differ less than 20\% in agreement to the pump-probe measurements. The synchronization phase for figs. \hyperref[fig:sync_sim]{\ref{fig:sync_sim}C(d-e)}
 is roughly $\phi=160^{\circ}$, also all the simulations for our system resulted in phase differences close to $\pi$, in agreement with the discussion in \cite{HolMeaMil1105} that the anti-phase synchronization is a more stable state when the oscillations amplitude $x_L,x_R$ are not identical.
 
The criteria for the optimum fitting is the matching of the laser frequency at which the bifurcation occurs, which is sufficient for explaining all of the non-linear phenomnena observed. The difference in the simulation power level and experimentally measured power level may be due to the variations in etched geometry, film thickness and optical losses. Fitting the entire spectra may provide a closer numerical match but it requires a full analysis taking account of of non-linear error propagation, detector and spectrum analyzer response function and multidimensional fitting that is beyond the scope of this paper.  

%\subsection{Phase noise}
%\begin{figure}[htbp!]
% \addtocounter{figureS}{1}
%  \centering
%  \includegraphics[width=0.6\textwidth]{noise.pdf}
%\vskip -0.1in
%\caption{Phase noise spectrum of the synchronized and individual oscillation states. CP: carrier power.}
%\label{fig:noisefig}
%\end{figure}
%Mutually synchronized oscillators can output oscillation signals that have lower phase noise than the output of individual oscillators. Such better noise performance is due to the suppression of low frequency fluctuations \cite{Kaka05,Hossein08} and the reduction of fundamental noise in correlated oscillators \cite{Chang97}. We increased the optical input power to $P_{in} = (35 \pm 3) ~$\textmu W to have sufficient carrier signal strength for the phase noise measurements. Fig.\ref{fig:noisefig} shows the phase noise curve for the individual oscillation states (red curve for $L$ OMO, blue curve for $R$ OMO) and the synchronized states (black curve). Indeed, we observe that the phase noise is lower in the synchronized state, particularly at $<$ 30 kHz frequency offset from the carrier. The drop in phase noise at low frequency offset may be due to a suppression of slow varying noise mechanisms in agreement with the previous observations \cite{Kaka05,Hossein08}. A complete phase noise analysis and stochastic study of synchronized oscillations are beyond the scope of the paper, since it requires a full treatment of coupled optomechanical dynamics, thermal noise and environmental perturbations.

%-----------transduction of mechanical motion------------ 

\end{document}